\documentclass{aa}

\usepackage{footnote}
\makesavenoteenv{tabular}
\makesavenoteenv{table}

\usepackage{hyperref}

\usepackage{booktabs}

\usepackage{subcaption,placeins,titlesec}%
\usepackage[labelfont=bf]{caption}
\usepackage{afterpage}
\usepackage{float}
\usepackage[flushleft]{threeparttable}
\usepackage{tikz,pgfplots}

\usepackage[arrows=pgf]{mhchem}

\usepackage{csvsimple}

\input{./packages.tex}
\newcommand{\twco}{\ce{^12CO}}
\newcommand{\thco}{\ce{^13CO}}
\newcommand{\cego}{\ce{C^18O}}
\newcommand{\hcn}{\ce{HCN}}

\newcommand{\hcop}{\ce{HCO+}}

\newcommand{\kms}{\ensuremath{\mathrm{km~s}^{-1}}}
\newcommand{\myr}{\ensuremath{\mathrm{Myr}}}
\newcommand{\erg}{\ensuremath{\mathrm{erg}}}
\newcommand{\persec}{\ensuremath{\mathrm{s}^{-1}}}
\newcommand{\mjykms}{\ensuremath{\mathrm{mJy~km~s}^{-1}}}
\newcommand{\mjy}{\ensuremath{\mathrm{mJy}}}
\newcommand{\jthrtwo}{~\ensuremath{{J=3-2}}}
\newcommand{\jfrthr}{~\ensuremath{{J=4-3}}}

\newcommand{\jtwoone}{~\ensuremath{{J=2-1}}}
\newcommand{\msun}{\ensuremath{\mathrm{M}_\odot}}
\newcommand{\mj}{\ensuremath{\mathrm{M}_\mathrm{J}}}
\newcommand{\lsun}{\ensuremath{\mathrm{L}_\odot}}
\newcommand{\mstar}{\ensuremath{M_*}}
\newcommand{\lstar}{\ensuremath{L_*}}
\newcommand{\mdisc}{\ensuremath{M_\mathrm{disc}}}
\newcommand{\mjup}{\ensuremath{M_\mathrm{Jup}}}
\newcommand{\zetacr}{\ensuremath{\zeta_\mathrm{CR}}}
\newcommand{\au}{\ensuremath{\mathrm{AU}}}
\newcommand{\av}{\ensuremath{\mathrm{A}_V}}
\newcommand{\micron}{\ensuremath{\mu\mathrm{m}}}
\newcommand{\teff}{\ensuremath{T_\mathrm{eff}}}
\newcommand{\kelvin}{\ensuremath{\mathrm{K}}}
\newcommand{\fuv}{\ensuremath{f_\mathrm{UV}}}
\newcommand{\luv}{\ensuremath{L_\mathrm{UV}}}
\newcommand{\puv}{\ensuremath{p_\mathrm{UV}}}
\newcommand{\tline}{\ensuremath{\tau_\mathrm{line}}}
\newcommand{\lxr}{\ensuremath{L_\mathrm{X-ray}}}
\newcommand{\rtaper}{\ensuremath{R_\mathrm{taper}}}
\newcommand{\rin}{\ensuremath{R_\mathrm{in}}}
\newcommand{\rout}{\ensuremath{R_\mathrm{out}}}
\newcommand{\amin}{\ensuremath{a_\mathrm{min}}}
\newcommand{\amax}{\ensuremath{a_\mathrm{max}}}
\newcommand{\apow}{\ensuremath{a_\mathrm{pow}}}
\newcommand{\vsep}{\ensuremath{v_\mathrm{sep}}}

\begin{document}

  \title{Thermochemical modelling of brown dwarf discs}

   \author{A.J. Greenwood$^1$
          \and
          I. Kamp$^1$ \and L.B.F.M. Waters$^{2,3}$ \and P. Woitke$^4$ \and W.-F. Thi$^5$ \and Ch. Rab$^6$ \and G. Aresu$^7$ \and M. Spaans$^1$
          }

   \institute{$^1$ Kapteyn Astronomical Institute, University of Groningen, Postbus 800, 9700 AV Groningen, The Netherlands\\
			  $^2$ SRON Netherlands Institute for Space Research, P.O. Box 800, 9700 AV Groningen, The Netherlands \\
			  $^3$ Anton Pannekoek Institute for Astronomy, University of Amsterdam, PO Box 94249, 1090 GE Amsterdam, The Netherlands \\
			  $^4$ SUPA, School of Physics \& Astronomy, University of St. Andrews, North Haugh, St. Andrews KY16 9SS, UK \\
			  $^5$ Max Planck Institute for Extraterrestrial Physics, Gie\textbeta{}enbachstra\textbeta{}e 1, 85741 Garching, Germany \\
			   $^6$ University of Vienna, Department of Astrophysics, T\"{u}rkenschanzstrasse 17, 1180 Vienna, Austria \\
			  $^7$ INAF, Osservatorio Astronomico di Cagliari, via della Scienza 5, 09047 Selargius, Italy\\
              \email{greenwood@astro.rug.nl}}

   \date{Received 24 July 2016 / Accepted 14 February 2017}

	\abstract{The physical properties of brown dwarf discs, in terms of their shapes and sizes, are still largely unexplored by observations. ALMA has by far the best capabilities to observe these discs in sub-mm CO lines and dust continuum, while also spatially resolving some discs. To what extent brown dwarf discs are similar to scaled-down T Tauri discs is currently unknown, and this work is a step towards establishing a relationship through the eventual modelling of future observations.

We use observations of the brown dwarf disc $\rho$ Oph 102 to infer a fiducial model around which we build a small grid of brown dwarf disc models, in order to model the CO, HCN, and \hcop{} line fluxes and the chemistry which drives their abundances. These are the first brown dwarf models to be published which relate detailed, 2D radiation thermochemical disc models to observational data.

We predict that moderately extended ALMA antenna configurations will spatially resolve CO line emission around brown dwarf discs, and that HCN and \hcop{} will be detectable in integrated flux, following our conclusion that the flux ratios of these molecules to CO emission are comparable to that of T Tauri discs. These molecules have not yet been observed in sub-mm wavelengths in a brown dwarf disc, yet they are crucial tracers of the warm surface-layer gas and of ionization in the outer parts of the disc.

We present the prediction that if the physical and chemical processes in brown dwarf discs are similar to those that occur in T Tauri discs -- as our models suggest -- then the same diagnostics that are used for T Tauri discs can be used for brown dwarf discs (such as HCN and \hcop{} lines that have not yet been observed in the sub-mm), and that these lines should be observable with ALMA. Through future observations, either confirmation (or refutation) of these ideas about brown dwarf disc chemistry will have strong implications for our understanding of disc chemistry, structure, and subsequent planet formation in brown dwarf discs.
}
	

	\maketitle

\section{Introduction}

{Brown dwarfs are very common objects in our Universe, yet they remain poorly understood because they are difficult to observe, classify, and model. Understanding brown dwarf protoplanetary discs is a crucial part of the overall picture of the formation of these objects, their later evolution, and the possibilities for planet formation. This paper focuses on the thermochemical modelling of brown dwarf protoplanetary discs.}

{Brown dwarfs are difficult to observe because of their very low luminosities. On the mass scale they straddle the boundary between massive planets and very low-mass stars. A brown dwarf is an object which is below the hydrogen-burning mass limit (about $0.08~\msun$) and thus cannot fuse hydrogen into helium  \citep{Oppenheimer:2000vo}. It has been that a lower mass limit for brown dwarfs should be defined as the threshold of deuterium fusion, at $0.013~\msun$. A less massive object where no nuclear fusion takes place, in the entire history of the object, is thus a planet \citep{Burrows:1995gr,Burrows:1997ua,Oppenheimer:2000vo}. The ``lithium test'' is a useful method to determine if a star is substellar: a main-sequence star will burn its lithium within $100~\myr$, while a brown dwarf may never reach the required temperature, leading to lithium enhancements in brown dwarfs \citep{Basri:1998tg}. Unfortunately, this discriminant is unreliable for objects that are much younger than the $\sim 100~\myr$ lifetime of lithium in a stellar atmosphere. Distinguishing and classifying brown dwarfs in young star-forming regions remains difficult, and often reliant on spectral type classification.}

{Survey data on brown dwarf discs are beginning to emerge which strongly suggest that the brown dwarfs in the nearby $\rho$ Ophiuchus star-forming region share a similar basic disc geometry with T Tauris -- that is, they are scaled-down T Tauri discs with {radii of the order of $20$ --  $150~\au$ \citep{Testi:2016tw}. However, because no unbiased sub-mm surveys of brown dwarf discs exist, the lower radius limit is unclear and there may also be larger brown dwarf discs that have yet to be resolved}. Brown dwarf and T Tauri discs also appear to have similar scale heights and degrees of flaring \citep{Guieu:2007dj,AlvesdeOliveira:2013df,Mohanty:2004io}. \cite{AlvesdeOliveira:2013df} and \cite{Mohanty:2013kl} suggest that brown dwarfs have disc masses of about $1\%$ of the brown dwarf mass -- a ratio that is similar to that of T Tauris. However, a \textit{Herschel} survey by \cite{Harvey:2012cz} shows that brown dwarf discs have a wide variety of disc masses, and that the ratio between disc mass and central object mass may instead be systematically lower in brown dwarf discs than for T Tauri discs.}

{On the contrary, from infrared \citep{Pascucci:2013te} and sub-mm \citep{Ricci:2014im,2012ApJ...761L..20R,Testi:2016tw} observational data, we expect that the discs may have up to several Earth masses of dust, which -- assuming a gas-to-dust ratio of 100 -- leads to total disc masses of a few Jupiter masses. Although the observed frequency of relatively massive brown dwarf discs is likely an observational bias, the presence of even a few fairly massive brown dwarf discs suggests that gas planets might form in some discs. It remains to be seen whether or not the planet formation process is efficient enough for this to happen. {The combination of low disc masses and a scarcity of giant planets around M dwarfs  suggests that most brown dwarf planets will be small and rocky \citep{Johnson:2010gu}}.}

{
Despite these uncertainties about disc masses and planet formation, the fact that brown dwarf discs appear broadly similar to a scaled-down T Tauri disc suggests that they may follow similar paths of evolution, and perhaps share similar formation scenarios. With further observational data in the mm wavelengths, there is a need for a contiguous set of  disc models which  describe the statistical distribution of brown dwarf discs.}

Although ALMA survey efforts are underway to analyse and resolve brown dwarf discs in both  dust continuum and \twco\jthrtwo{} emission, there is likely a large population of brown dwarf discs that cannot be spatially resolved, even with ALMA. {Brown dwarf discs are thought to be much smaller than T Tauri discs, where} numerical simulations have shown that a majority of brown dwarf discs have a radius of $\approx 10~\mathrm{AU}$ \citep{Bate:2003cv,Bate:2009br,Bate:2012hy}. The likelihood that many brown dwarf discs are small is beginning to be backed up with observational data: in the ALMA survey of \cite{Testi:2016tw}, 8 out of the 11 detected sources in Ophiuchus have unresolved radii $ \rout \lesssim 24~\au$.  Similar work by 	\cite{vanderPlas:2016do} surveys 7 Upper Scorpius OB1 brown dwarfs and one Ophiuchus brown dwarf using ALMA, where all sources remain unresolved (suggesting discs of $\rout \lesssim 40~\au$). A few brown dwarf discs are known to be very large: \cite{Ricci:2014im} find three discs that have large outer radii of about $70~\au$, $140~\au$, and $>80~\au$, detected with ALMA in the 0.89 mm and 3.2 mm dust continua. However, these are likely to be outliers and not representative of the general population. Thus even if they are difficult to observe, the more compact (but much more populous) regime of brown dwarf discs should not be forgotten.

Currently there are no known planets orbiting a solitary brown dwarf, barring systems that have doubtful formation scenarios or whose host is likely not to be substellar. For example, \cite{Han:2013by} find through microlensing observations a $1.9 \pm 0.2~\mj$ planet orbiting a $0.022~\msun$ host, at a separation of \SI{0.87}{\au}. The mass ratio of $0.08$ is very high, and there is no clear explanation for how this system formed. Perhaps the closest system yet to satisfying this ongoing search is TRAPPIST-1 \citep{Gillon:2016hl}, where a few planets of a few Earth masses are found transiting a very cool dwarf star. Although they conclude that the star \emph{most likely} exists on the main sequence, it has not been ruled out that the star is a brown dwarf. This system presents very strong evidence that planets may form in discs around stars that sit on the substellar boundary. Through both thermochemical modelling and observations, we can gain a better understanding of the structure and evolution of brown dwarf discs -- leading also towards understanding the formation and distribution of planets around brown dwarfs.

{In this paper, we use the thermochemical disc modelling code ProDiMo \citep{Woitke:2009jf,Kamp:2010ek,Aresu:2011cm} to produce a small grid of disc models in order to illustrate the effects of disc radius on the line fluxes of key molecules in the sub-mm regime, the distribution of these molecules in the disc, to check which reactions dominate their formation and destruction, and to compare these properties to T Tauri models.} We also use the grid to predict line fluxes, in preparation for ALMA observations of brown dwarf discs. ProDiMo is a well-established code which has been calibrated against observations -- for example, \cite{Woitke:2016gp} is the first in a series of papers which will describe the standardization of these models.\footnote{Our grid has been developed from the DIANA framework of models, under the European FP7 project DiscAnalysis. See \href{http://www.diana-project.com}{See http://www.diana-project.com} for more details.}

The premise of this paper is to bring together both observations and advanced thermochemical models of brown dwarf discs for  the first time. We produce a small grid of disc models, based around a fiducial model {which fits the sub-mm continuum and (within the errors) the CO line observations of the $\rho$ Oph 102 disc.} The models serve to show that HCN, \hcop{}, and some CO isotopologues may be detectable in brown dwarf discs using ALMA, paving the way for future observational surveys that probe the composition of these discs.

\section{Models}
\label{sec:models}
The ProDiMo code uses a fixed set of disc structure and stellar parameters. Using a 1+1D {fixed} disc structure, the 2D dust continuum radiative transfer solution is computed for the entire disc using a ray-based method \citep{Woitke:2009jf}.
The chemistry and gas heating / cooling balance routines solve for chemical and thermal equilibrium, based on the local 2D continuum radiation field. The last computation stage is the gas line radiative transfer, which produces data cubes of the line profiles of common species such as \twco{}, \hcn{}, and \hcop{} \citep{Woitke:2009jf}.

The dust grain size distribution used in the models follows a power law, with the inclusion of \SI{}{\mm}--sized dust grains; this is typical in modelling T Tauri discs. We follow the prescription given in \cite{Woitke:2016gp}, and assume that it also holds for a brown dwarf disc. Indeed there is strong evidence for such large dust grains in the disc of $\rho$ Oph 102, where it is shown by  \cite{2012ApJ...761L..20R} that the opacity index $\beta$ must be smaller than $1$ if the disc radius is greater than \SI{5}{\au}.

{The density of the discs follows a power law, with an exponential tapering-off starting at the radius \rtaper{}. The disc model extends to \rout{}, and \rtaper{} is fixed at $\rtaper =  0.125 \times \rout$. This scaling relation has been chosen because it allows the outer edge of the fiducial model to reach a column density of $N_{<\mathrm{H}>} \approx 10^{20}~\mathrm{cm}^{-2}$, where  $N_{<\mathrm{H}>} = N_\mathrm{H} + 2N_{\mathrm{H}_2}$. This is justified by \cite{Woitke:2016gp}, who argue that the outer radius should be large enough such that even the \twco{} $\mathrm{J}=1-0$ line becomes optically thin, and suggest setting $N_{<\mathrm{H}>}(\rout) \approx 10^{20}~\mathrm{cm}^{-2}$ as the criterion. The smaller ($\rout \leq 40~\au$) discs in the grid truncate at up to $N_{<\mathrm{H}>} \approx 10^{21}~\mathrm{cm}^{-2}$, while the larger ($\rout \geq 100~\au$) discs truncate at densities as low as $N_{<\mathrm{H}>} \approx 10^{18}~\mathrm{cm}^{-2}$. The relatively sharp truncation of the smaller discs is supported by ALMA data, which suggest that most brown dwarf discs in the Ophiuchus region are dynamically truncated to small radii ($\rout \lesssim 25~\au$, \citealp{Testi:2016tw}). }

\subsection{Defining the fiducial model}

In order to define a ``starting point'' for our grid, {we choose $\rho$ Oph 102 as an object to anchor our small model series around, so that our input parameters are reasonable.} $\rho$ Oph 102 is a comparatively well-studied brown dwarf disc. ALMA has detected a \twco \jthrtwo{} line flux of $530 \pm 45 ~ \mjykms$, with unresolved dust continuum observations constraining the radius of the mm dust disc to $\rout \lesssim 40~\au$ \citep{2012ApJ...761L..20R}. They also find that the observed \twco{} \jthrtwo{} flux cannot feasibly be reproduced with a  very small, optically-thick disc. Thus, their model of the CO flux places a lower limit of about \SI{15}{\au} on the outer radius of the disc. Although it is unclear if the disc is \emph{representative} of the general low-mass disc population, it is a well-studied candidate to adopt as a fiducial reference. The large, resolved discs of \cite{Ricci:2014im} are relatively bright in CO but are not likely to be prototypical examples of brown dwarf discs.

\begin{figure}[tbh]
\centering
\includegraphics[width=0.42\textwidth]{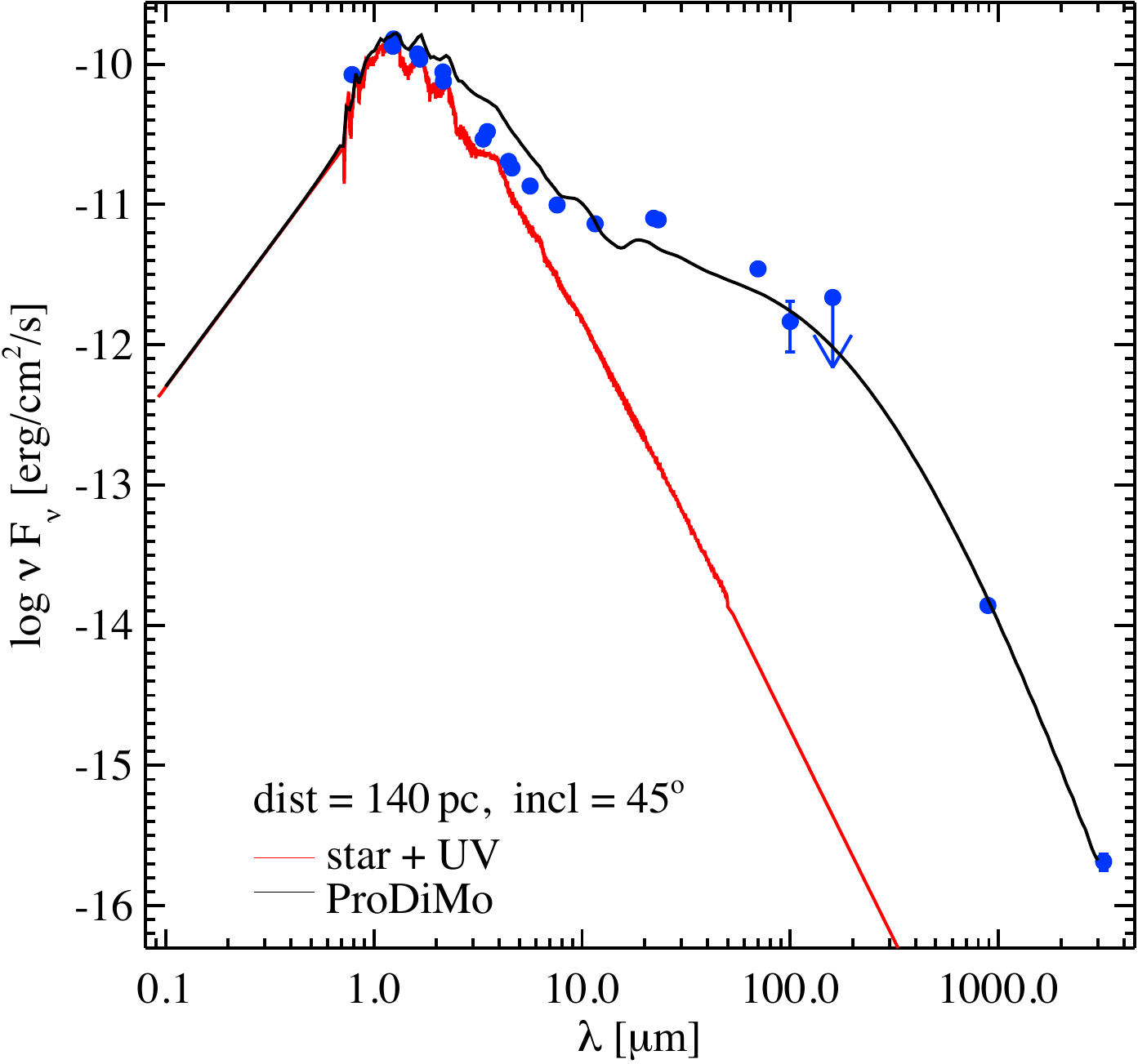}
\caption{Fit of the fiducial model (black line) to the observed disc SED (blue points), from the data in Table \ref{tab:seddata}. The red line is the synthetic stellar spectrum (plus UV excess). A downwards arrow marks an upper limit datum from \textit{Herschel} PACS at 160 \micron{}.}
\label{fig:oph102sed}
\end{figure}

In order to begin modelling the disc, we must define a model photosphere. We compute the photospheric fit ourselves, rather than taking literature values, to ensure consistency with past and future ProDiMo models that use the same fitting routine and synthetic photospheres. Using the known photometry, we fit the photosphere with a luminosity of $0.0822~\lsun$ and reddening $E_{B-V}=1.5815$ (where $R_V=3.1$), and an effective temperature of $3000~\kelvin$ (see Fig. \ref{fig:oph102sed}). Some chemistry processes are highly sensitive to the properties of the stellar spectrum:  it is important to treat this as accurately as possible. An evolutionary strategy was used in order to fit the photosphere (see \citealp{Woitke:2011fo}), using Drift-Phoenix brown dwarf synthetic spectra\footnote{
Drift-Phoenix \citep{Witte:2009br,Witte:2011kn} is a code which couples the Phoenix model atmosphere code \citep{Hauschildt:1999vw,Baron:2003tk} with the Drift code which can model cloud formation in substellar objects \citep{Woitke:2003cs,Woitke:2004ie,Helling:2006gp}.
}. There remains some degeneracy in the fit due to uncertain levels of reddening and extinction, but there is a lack of consensus in previous literature values. \cite{Natta:2002ea} report $\av=3.0$, \cite{Muzic:2012jo} report $\av=3.7$ from $J-K$ colour and $\av=6.0$ from spectral fitting, and \cite{Manara:2015ey} report $\av=2.2$. We note that \citeauthor{Muzic:2012jo} refer to $\rho$ Oph 102 as GY204. Our fitted photosphere falls within a few hundred Kelvin of previous estimates, but we cannot ascertain the accuracy of this because the visual extinction is so uncertain. 

Simultaneous photometry and accurate extinction measurements are needed to better characterize the photosphere, as the true level of reddening has a significant influence on the modelled luminosity (which then directly influences the disc chemistry and line fluxes). Even given better data, the task of determining accurate stellar parameters is difficult because the theory of brown dwarf atmospheres is incomplete \citep{Helling:2008hp,Helling:2014fx}. Accretion is another significant factor, as strong levels of accretion increase the irradiation of the disc by X-rays, and variable accretion can lead to differing observations.

The parameter space which ProDiMo is capable of exploring is very large: it is possible to create numerous considerably different models which still fit the few observational data that exist. Thus, we adopt and adapt the disc parameters from those which are typical of T Tauri discs (see Sect. \ref{sec:fiducialmodel}). The dust parameters of \apow{}, \amin{} and \amax{} in Table \ref{tab:gridparameters} were found through a coarse Monte Carlo fitting procedure. This follows the method of \cite{Tilling:2012cj} to fit the spectral energy distribution (SED) of $\rho$ Oph 102, and is consistent with the expectation that there are mm-sized dust grains in the disc \citep{2012ApJ...761L..20R}. The SED fit alone is highly degenerate, but it is nevertheless encouraging that a standardized model can provide a reasonable fit to the data in both dust continuum and CO line fluxes.

  \begin{table}[!htb]
    \sisetup{round-mode=places}
    \caption{SED data for $\rho$ Oph 102, collated by the VOSA service \citep{Bayo:2008bk}. The 160 \micron{} PACS data point is an upper limit.}
    \label{tab:seddata}
    \centering
	
    \begin{tabular}{
      S[round-mode=figures,round-precision=3]
      S[round-mode=figures,round-precision=3]
      S[round-mode=figures,round-precision=3]
 r  
    }
{Wavelength} & {F$_\nu$ (mJy)} & {$\sigma$ (mJy)}        & {Source}                         \\
{(\micron)} \\ \midrule
    0.7862101596600001 & 1.68805025414070  & 0.0777375870282000    & DENIS.I $^1$       				  \\
    1.2210602758100000 & 16.7857386361489  & 1.08221656363770      & DENIS.J $^1$                      \\
    1.2350000000000001 & 16.9324099808586  & 0.374287822179700     & 2MASS.J $^2$                      \\
    1.2443966151400001 & 19.5226073588536  & 0.0172848825720000    & UKIDSS.J $^3$                     \\
    1.6221246869299999 & 29.8045267609795  & 0.0159300775533000    & UKIDSS.H $^3$                     \\
    1.6620000000000001 & 28.8317660845307  & 0.637320909992900     & 2MASS.H $^2$                      \\
    2.1465009653000000 & 37.8410257379474  & 2.43970108949840      & DENIS.KS $^2$                     \\
    2.1590000000000003 & 32.8876359322127  & 0.636103274027200     & 2MASS.KS $^2$                     \\
    3.3526000000000002 & 24.5147267444518  & 0.541893545861700     & WISE.W1 $^4$                      \\
    3.5075111435999999 & 29.4000001961356  & 1.41000000940650      & \textit{Spitzer} IRAC3.6 $^5$     \\
    4.4365778685099997 & 24.2999991621123  & 1.18000000787210      & \textit{Spitzer} IRAC4.5 $^5$     \\
    4.6028000000000002 & 22.8528643104180  & 0.442013583424400     & WISE.W2 $^4$                      \\
    5.6281016868600009 & 21.6000001440999  & 1.04000000693810      & \textit{Spitzer} IRAC5.8 $^5$     \\
    7.5891586020100004 & 22.2999991487702  & 1.10000000733840      & \textit{Spitzer} IRAC8.0 $^5$     \\
   11.5608000000000004 & 26.0061557498146  & 0.646718974800500     & WISE.W3 $^4$                      \\
   22.0883000000000003 & 56.4196251019124  & 3.42965007559610      & WISE.W4 $^4$                      \\
   23.2096043690170006 & 57.9000023862667  & 5.36000003575810      & \textit{Spitzer} MIPS.24 $^5$     \\
   70.0000000000000000 & 80.1000000000000  & 5.60000000000000      & \textit{Herschel} PACS70 $^6$     \\
  100.0000000000000000 & 48.4000000000000  & 19.1000000000000      & \textit{Herschel} PACS100 $^6$    \\
  160.0000000000000000 & {$\leq$}115.200000000000  & 55.2000000000000      & \textit{Herschel} PACS160 $^6$  \\
  890.0000000000000000 & 4.10000000000000  & 0.220000000000000     & ALMA $^7$                   \\
 3200.0000000000000000 & 0.22000000000000  & 0.0300000000000000    & ALMA $^7$                                              \\ \midrule
    \end{tabular}
\raggedright
References are as follows: 1: \cite{Consortium:2005vp}, 2: \cite{Skrutskie:2006hl}, 3: \cite{Lawrence:2007hu}, 4: \cite{Ochsenbein:2000fm}, 5: \cite{Evans:2003bo,Evans:2009bk}, 6: \cite{AlvesdeOliveira:2013df}, 7: \cite{2012ApJ...761L..20R}.

\end{table}

\subsection{\twco{} sub-mm lines from the fiducial model}

\begin{figure}[tbh]
\centering
\includegraphics[width=\linewidth]{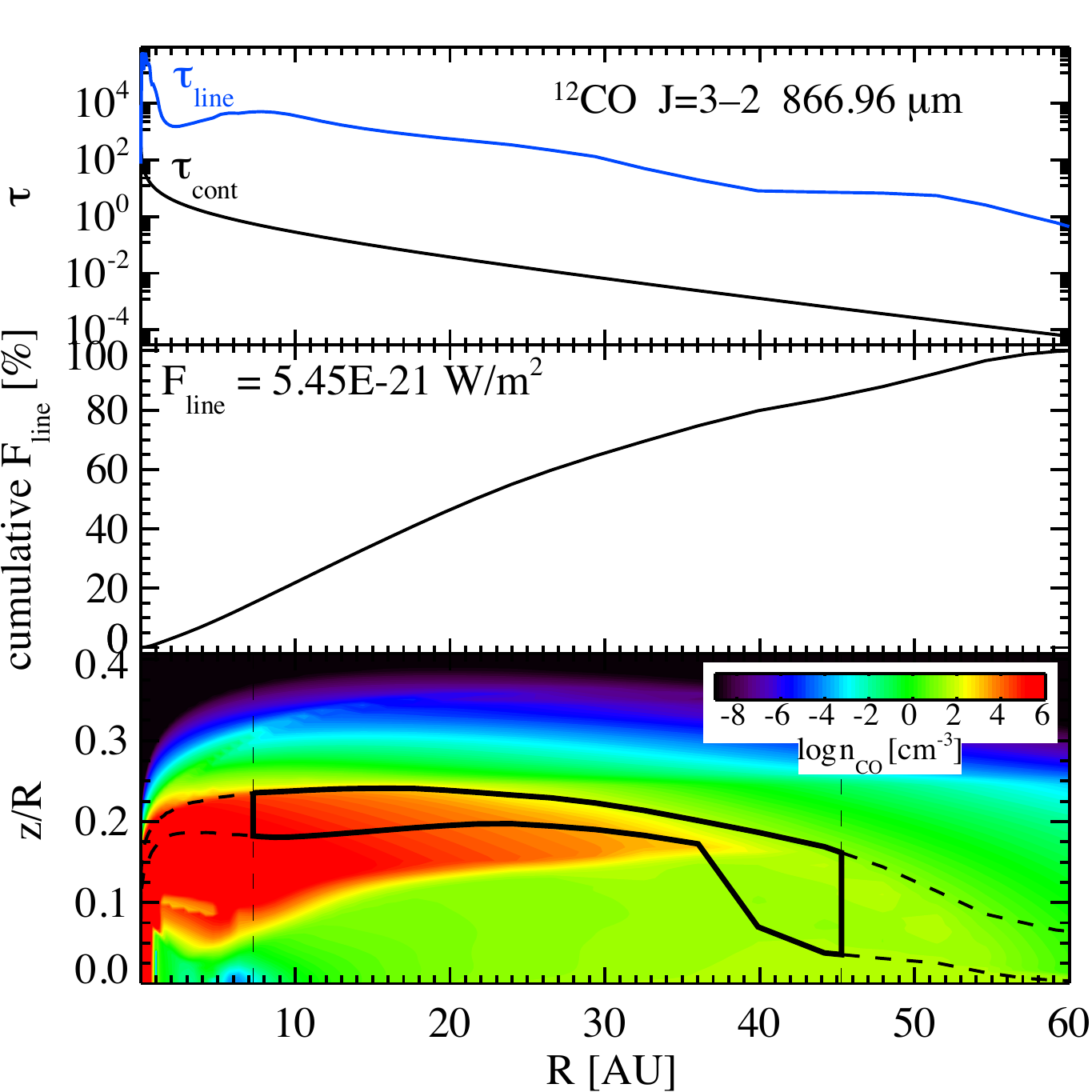}
\caption{Analysis of the \twco{} \jthrtwo{} line in the fiducial model. As seen in the top panel, most of the line emission remains optically thick. The middle panel shows the cumulative line flux at each radius.  In the lower panel, the area enclosed by the black curve shows where $70\%$ of the flux originates from in both the radial and vertical directions (hence, the enclosed area contains $70\% \times 70\% = 49\%$ of the total line flux). {The x-axis boundaries of this area are created first, as the radii within which 15\% and 85\% of the radially-cumulative line flux originates. Then the y-axis boundaries of this area are taken as the disc heights $(z/R)$, where within each radial sector 15\% and 85\% of the vertically-integrated line flux originates.}}
\label{fig:twcodist}
\end{figure}

In addition to the SED, we can also compare the disc models with the \twco{} \jthrtwo{} line flux of $\rho$ Oph 102. \cite{2012ApJ...761L..20R} report a line flux of $530 \pm 45 ~ \mjykms$, while our closest model in the grid (see Sect. \ref{sec:fiducialmodel}), which we choose as the fiducial model, under-reports this slightly at $416~\mjykms$. Some simple explanations for the discrepancy could be an inaccurate stellar luminosity, or that the disc is more massive than $4\times 10^{-4}~\msun$. Figure \ref{fig:twcodist} shows that $\sim 80\%$ of the \twco{} \jthrtwo{} line flux builds up within $40~\au$, which is consistent with the very tentative indications of extended emission by \cite{2012ApJ...761L..20R} which suggest that the CO emission extends beyond this radius. Higher resolution observations  show that the dust disc is likely $\lesssim 24~\au$ in radius \citep{Testi:2016tw}. {The gas and dust around higher mass objects are not always co-spatial (for example, see \citealp{Qi:2003dn}). Thus, it seems reasonable to suggest that that the dust and gas in the disc around $\rho$ Oph 102 may similarly be non-co-spatial}.

\cite{2012ApJ...761L..20R} also report that the CO gas in $\rho$ Oph 102 spans a range of channel map radial velocities of the order of $1~\kms$ (but no peak separation is reported), which is consistent with the $1.19~\kms$ \twco{} \jthrtwo{} peak separation of our fiducial model. However, there is still significant uncertainty as only an upper limit on the inclination of the disc $(i \leq 80^\circ)$ is known. Our fiducial model has an inclination $i=45^\circ$. The strongest statement that we can make about $\rho$ Oph 102 here is that there is no indication that its true radius differs significantly from what we assume. Better observational data are needed to strongly constrain the disc inclination.

\subsection{The fiducial model and grid selection}
\label{sec:fiducialmodel}

There is currently little general consensus on what a ``normal'' brown dwarf disc looks like. The numerical simulations by {\cite{Bate:2003cv} and \cite{Bate:2009br,Bate:2012hy} suggest that most brown dwarf discs are quite small: $95\%$ of brown dwarf discs may have radii $\rout \lesssim 10~\au$ \citep{Bate:2003cv}}. However, the size distribution is still unconfirmed through observations. The only resolved objects are that of \cite{Ricci:2014im} and \cite{Testi:2016tw}.

The grid is based upon the properties that we could reasonably expect to see in a scaled-down version of a ``typical'' T Tauri disc. Table \ref{tab:gridparameters} describes some of the model parameters, that are based around the fiducial model (which coarsely fits $\rho$ Oph 102) and similar to the reference T Tauri models proposed by \citep{Woitke:2016gp}. The small model grid is focused on exploring how the disc radius may affect gas line observations of brown dwarf discs from mm and sub-mm observatories such as ALMA. We chose a grid that varies the disc mass, taper radii fixed at one-eighth of the (varying) outer radius, and all other parameters remaining constant. The disc parameters (such as scale height, dust grain size distribution, and dust-to-gas ratio) are consistent with what a scaled-down version of a T Tauri disc that has millimetre-sized dust grains and an M-type star could look like.

We also model the reference T Tauri disc, using the parameters in Table \ref{tab:gridparameters} and same version of the ProDiMo code as the rest of our models. This ensures consistency so we can later compare the T Tauri model with the brown dwarf models.

\begin{figure}[tbh]
\centering
\includegraphics[width=0.45\textwidth]{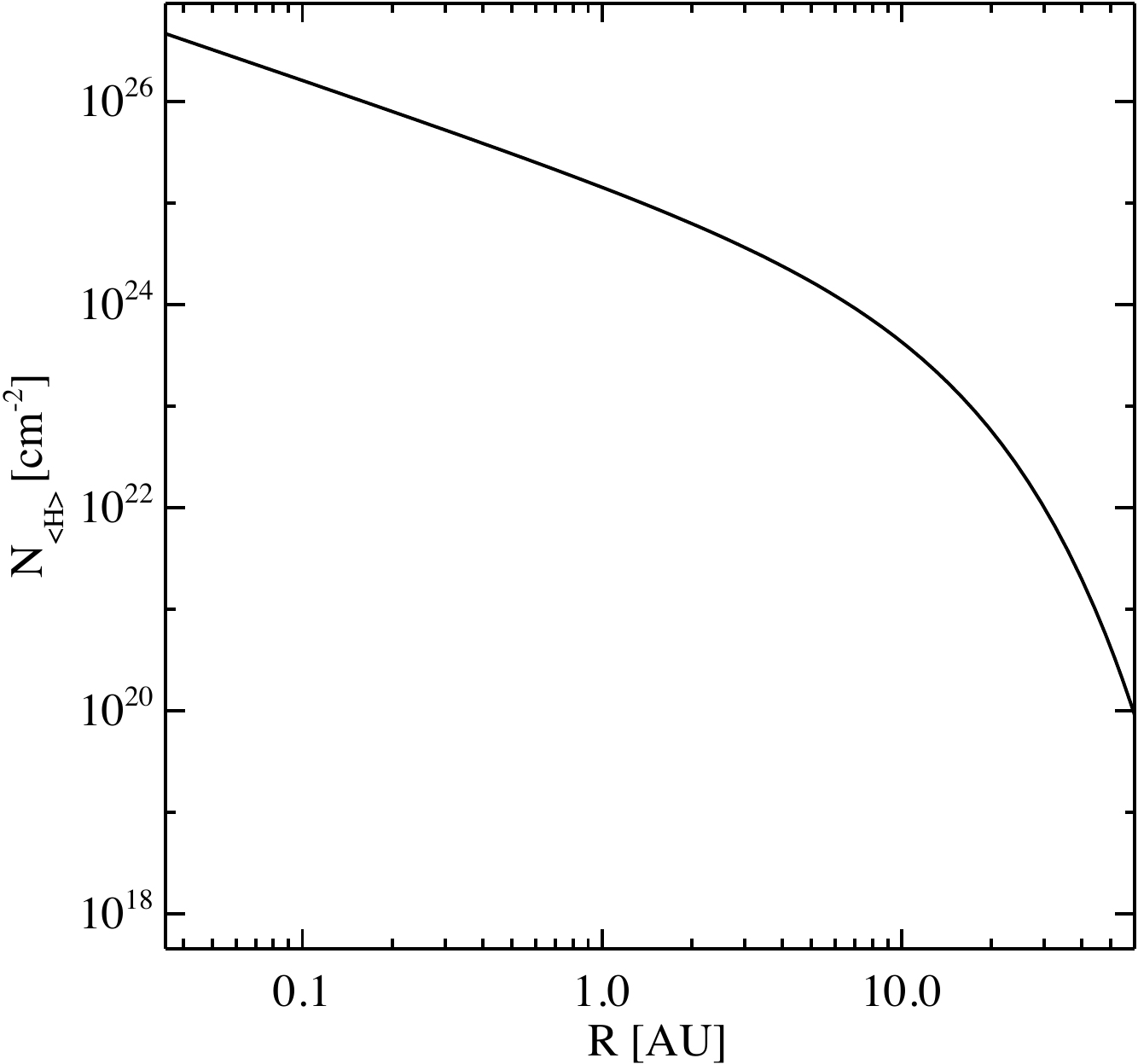}
\caption{Total hydrogen column density for the $\mdisc = 4 \times 10^{-4} ~\msun$, $\rtaper = 7.5 ~\au$ fiducial model, illustrating that the column density of the disc begins to decrease exponentially at \rtaper, with exponent $-1.0$. {As discussed in Sect. \ref{sec:models}, we set the taper radius so that $N_{<\mathrm{H}>} \approx 20~\mathrm{cm}^{-2}$ at the outer edge of the fiducial model.}}
\label{fig:surfdens}
\end{figure}

\begin{figure*}[tbh]
\centering
\includegraphics[width=0.32\linewidth]{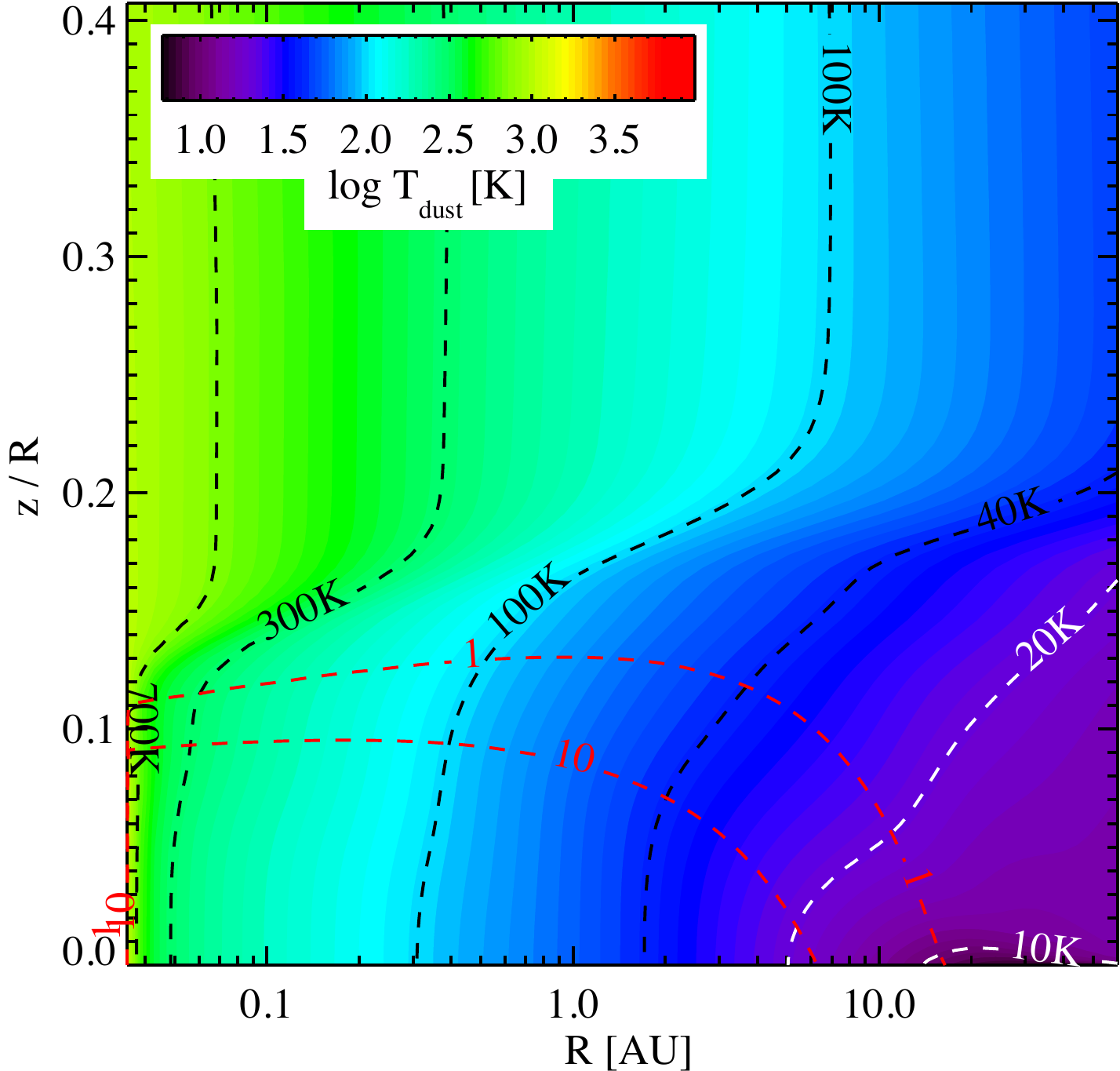}
\hspace{0.5em}
\includegraphics[width=0.32\linewidth]{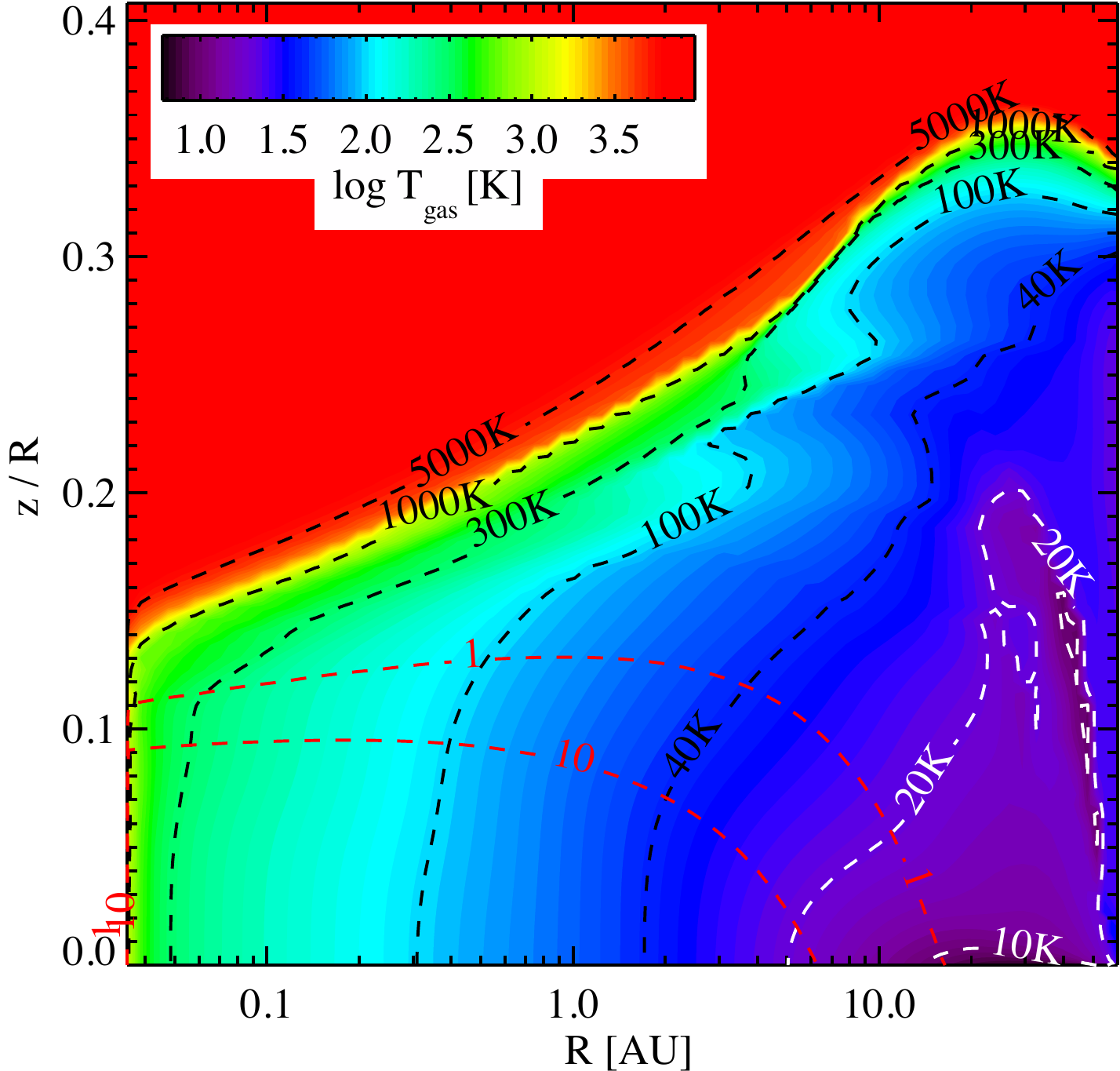}
\hspace{0.5em}
\includegraphics[width=0.32\linewidth]{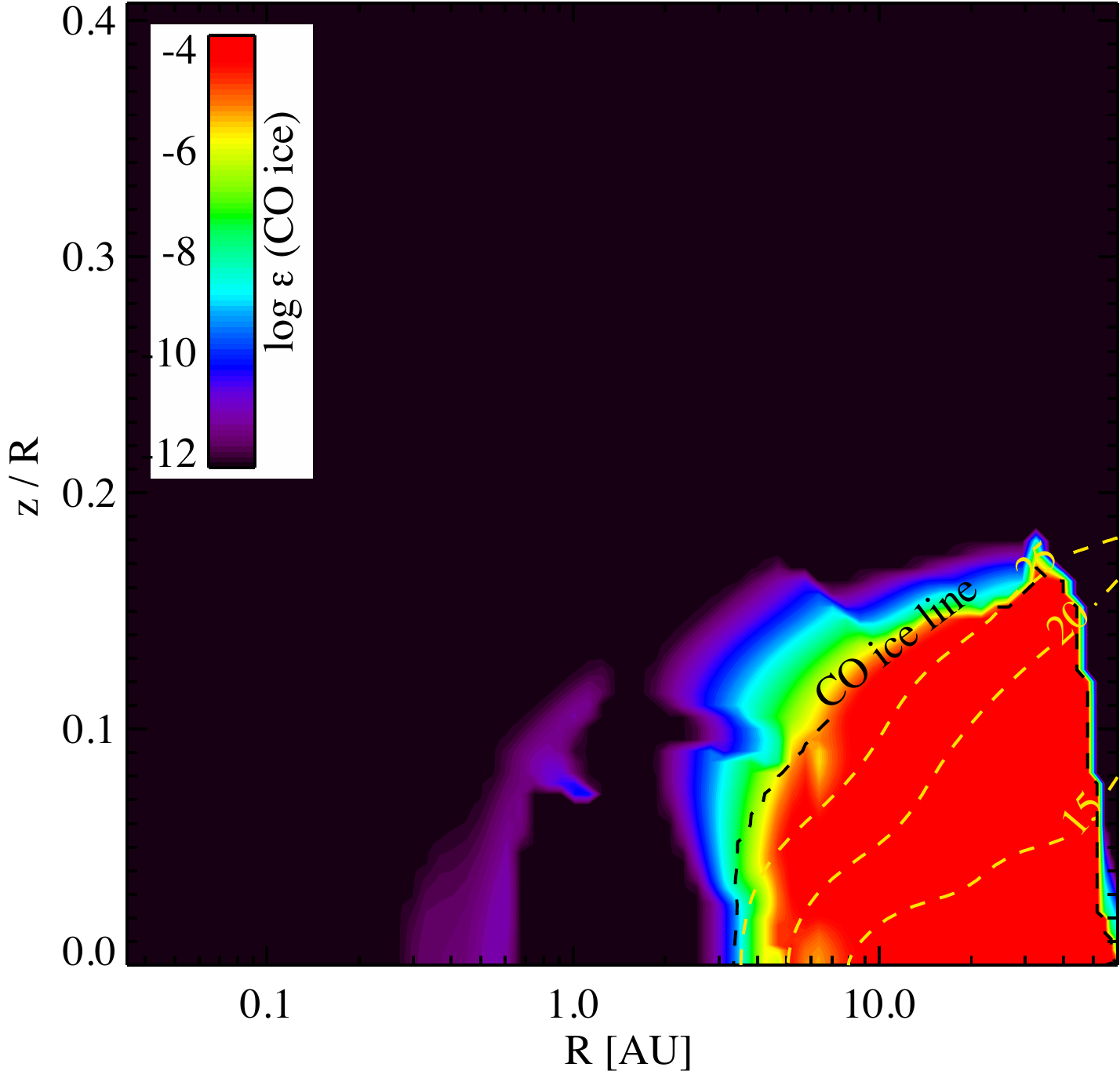}
\caption{\textbf{Left:} dust temperature structure of the fiducial model. Overlaid on this plot (and the middle plot) are contours for the plotted temperatures, and extinction contours for vertical $\av=1$ and $\av=10$. \textbf{Middle:} gas temperature structure of the fiducial model. \textbf{Right: } CO ice line. We follow the ice line definition of \citep{stefanothesis}, with the ice line being defined as where the gas and ice abundances are equal. Dust temperature contours are marked at 15, 20, and $25~\kelvin$.}
\label{fig:coice}
\end{figure*}

The fiducial model is described by {Figs.} \ref{fig:surfdens} and \ref{fig:coice}, where the dust and gas temperatures, CO ice line, and hydrogen column density are shown. The basic  structures of the disc appear very similar to the ``standardized'' T Tauri ProDiMo models (\citealp{Woitke:2016gp}, Rab et al. in prep.). That is, if the T Tauri models were scaled in radius then the chemical structure between the two types of model is self-similar. For example, such a T Tauri disc has an optically-thick midplane ($\av > 10$) out to about $30~\au$, compared to $6~\au$ in the fiducial model. When ignoring the scale of the radius axis, the two discs look remarkably similar in basic properties such as the gas and temperature structure and ice line locations.

\begin{table*}[tb]
\begin{center}
\caption[]{Fundamental parameters of the model grid, compared to that of the reference T Tauri model by \cite{Woitke:2016gp}. Every possible combination of \rout{} and \mdisc{} is used, to create a grid of 50 models. Parameter definitions are further explained by \cite{Woitke:2009jf}.}
\label{tab:gridparameters}
\medskip
\begin{tabular}{l l r r}
Symbol & Quantity (units) & Grid value(s) & \parbox[t]{5cm}{\raggedleft Reference T Tauri (parameters from \citealt{Woitke:2016gp})} \\ \midrule
$\mstar$ & Stellar mass (\msun) & $0.06$ & $0.7$ \\
$\lstar$ & Stellar luminosity (\lsun) & $0.0822$ & $1$ \\
\teff & Effective temperature (\kelvin) & $3000$ & $4000$ \\
$\fuv$ & UV excess ($\luv / \lstar$) & $0.01$ & $0.01$ \\
$\puv$ & UV power law exponent & $1.0$ & $1.3$ \vspace{0.13em} \\
$\lxr$ & \parbox[t]{6cm}{X-ray luminosity (\erg~\persec, bremsstrahlung continuum) $^1$} & $10^{29}$  & $10^{30}$ \vspace{0.13em} \\
$\zetacr$ & Cosmic ray \ce{H2} ionization rate ($\persec$) & $1.3 \times 10^{-17}$ & $1.7 \times 10^{-17}$ \\
$\mdisc$ & Disc mass ($\times 10^{-4}~\msun$) & $1$, $2$, $4$, $8$, $15$ & $100$\\
${\rho_\mathrm{d}} / {\rho_\mathrm{g}}$ & Dust-to-gas ratio & $0.01$ & $0.01$ \\
$\rin$ & Inner disc radius (\au) & $0.035$ & $0.07$ \vspace{0.13em} \\
$\rout$ & Outer disc radius (\au) & \parbox[t]{2.9cm}{\raggedleft $20$, $40$, $60$, $80$, $120$, $160$, $200$, $400$, $600$, $800$} & $600$ \vspace{0.13em} \\
$\rtaper$ & Tapering-off radius (\au) & $0.125\times\rout$ & $100$\\
$H_0$ & Scale height at $100~\au$ (\au) & $10$ & $10$ \\
$\beta$ & Flaring exponent $H(r)=H_0 \left( {r} / {r_0} \right)^\beta $ & $1.15$ & $1.15$ \vspace{0.13em} \\
$N$ & Number of grid points & $80 \times 80$ & $160 \times 150$ \vspace{0.13em} \\
$\apow$ & Dust size distribution $f(a)\propto a^{-\apow}$ & 3.5 & 3.5 \\
& \multicolumn{3}{l}{\parbox[t]{6cm}{Dust grain mixture: $60\%$ amorphous \ce{Mg_{0.7} Fe_{0.3} Si O_3} silicates $^2$, $15\%$ amorphous carbon $^3$, $25\%$ vacuum for porosity~$^\dag$}} \vspace{0.13em} \\
$\amin$ & Min. dust grain size (\micron) & $0.05$ & 0.05\\
$\amax$ & Max. dust grain size (\micron) & $3000$ & 3000\\
$i$ & Inclination angle ($^\circ$) & $45$ & $45$ \vspace{0.13em} \\
$\alpha$ & \parbox[t]{6cm}{Turbulent viscosity, for Dubrulle settling of dust grains $^4$} & $10^{-3}$ & $10^{-2}$ \vspace{0.13em} \\
$\chi_{\mathrm{ISM}}$ & \parbox[t]{6cm}{Strength of incident UV w.r.t. ISM field $^5$} & 1 & 1 \\
\midrule
\end{tabular}
\end{center}
\raggedright
References are as follows: 1: \cite{Woitke:2016gp}, 2: \cite{Dorschner:1995wq}, 3: \cite{Zubko:1996fn}, 4: \cite{Dubrulle:1995jn}, 5: \cite{Draine:1978ec}.
\dag: {The dust is a distribution of hollow spheres, where the maximum fractional volume filled by the central void is 0.8 \citep{Min:2005uy,Min:2016hr}.}
\end{table*}

\subsection{{Chemical modelling and comparisons with T Tauri discs}}
{
Here we outline some of the details of the chemical network and modelling procedures, but a detailed discussion of the chemical networks in ProDiMo is contained in a forthcoming paper (Kamp et al., in prep.).}

\begin{figure}[tbh]
\centering
\includegraphics[width=\linewidth]{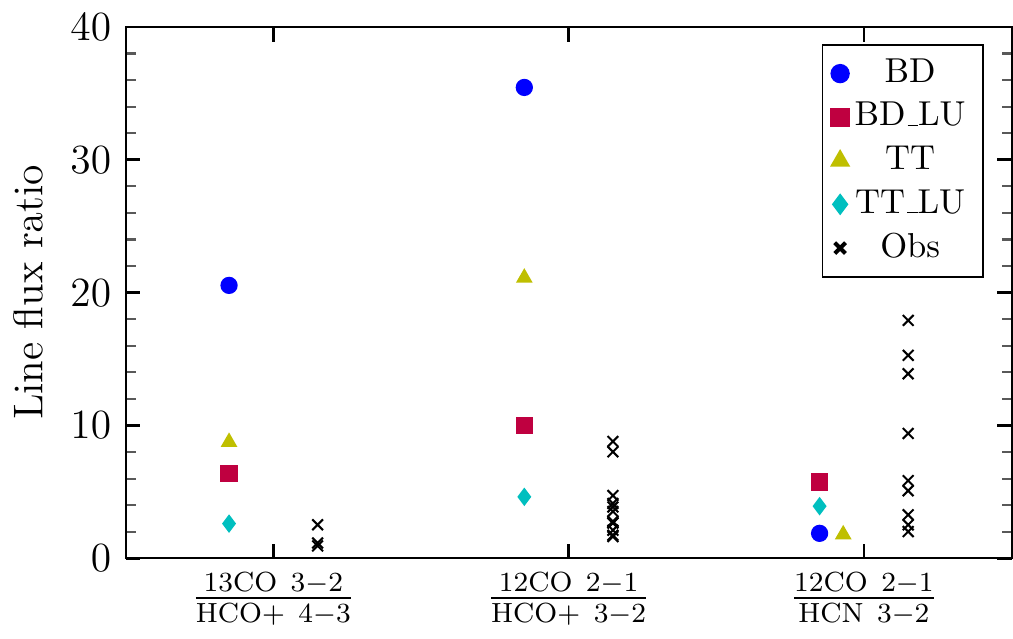}
\caption{Line flux ratios of our fiducial model (BD) against that of a selection of observed discs around T Tauri and Herbig objects (Obs), our T Tauri model (TT), and our models with large chemical networks (BD\_LU and TT\_LU). The \thco\jthrtwo{} to \hcop{}\jfrthr{} ratio data are from \cite{Salter:2011jn}. The data for the other two line ratios are from \cite{Oberg:2010kf,Oberg:2011ew}.}
\label{fig:linefluxratios}
\end{figure}

{Our model grid uses a small and computationally-efficient chemical network. For comparison, we also re-ran the fiducial model with a large and more computationally expensive network.} {The more complex network increases the number of species from 100 species to 235 species (Kamp et al., in prep.), using reactions from the UMIST 2012 database \citep{McElroy:2013ki}}. {The small 100-species network includes freeze-out of CO, \ce{H2O}, \ce{CO2}, \ce{CH4}, \ce{NH3}, \ce{SiO}, \ce{SO2}, \ce{O2}, \ce{HCN}, and \ce{N2}. The large 235-species network includes freeze-out of all neutral species except noble gases (Kamp et al., in prep.).} The small 100-species network models appear to underpredict \hcop{} fluxes by a factor of a few, and to overpredict the HCN fluxes by a factor of a few. The reasons for this are discussed in Sect. \ref{sec:lineratiosandfluxes}, but they pertain to the fact that the small network does not include all species (and thus reactions) that are significant to the formation and destruction of these sensitive molecules. In Fig. \ref{fig:linefluxratios} we show the modelled line ratios for common molecules in our fiducial model and compare these both to observations, and to models with a larger and more complex chemical network.

{In all of our models, water can form on grain surfaces by the Eley-Rideal mechanism (\citealt{Hollenbach:2008ho}, ProDiMo implementation described by \citealt{Kamp:2013ix}), which is important at intermediate heights above the water freeze-out zone. All of our models use time-dependent chemistry (to an age of 3 Myr), with reactions and adsorption energies ($E_{\mathrm{ads}}$), which are important for freeze-out) from the UMIST 2012 database \citep{McElroy:2013ki}. Further details of the reactions included can be found in \cite{Woitke:2016gp}.}

Rab et al. (in prep.) find that for T Tauri discs, ProDiMo models must be metal-depleted (relative to solar abundances) in order to accurately predict the fluxes of molecules such as \hcop{} and \hcn{}. The prescription is that all elements except H, He, C, N, O, Ne, and Ar are depleted by a factor of 100 relative to the $\zeta$ Oph diffuse cloud \citep{Graedel:1982hf}. Every model in this paper follows the same depletion prescription.

{
There is relatively little prior work to compare this work with. \cite{Wiebe:2008dy} published some of the first models comparing the chemistry of T Tauri and brown dwarf discs. However, their $1+1~\mathrm{D}$ models explore a very different parameter space and focus on the chemical evolution of the disc, thus they are not easily comparable. Their disc models have different surface density profiles, have somewhat larger disc masses of 2.2 to $5.7~\mjup$, assume that the brown dwarf discs extend from $\rin=0.03~\au$ to a very large $\rout=800~\au$, and do not report any line or column density ratios.}

{\cite{Walsh:2015jr} model an M dwarf disc, a T Tauri disc, and a Herbig Ae disc, and compare their molecular compositions. Similarly to \cite{Wiebe:2008dy}, \rin{} is the same for all models. However, the dust condensation radius depends directly upon the stellar parameters. \rout{} is arbitrarily large as they focus on the inner $10~\au$ of the disc. In contrast, our ProDiMo models focus on the sub-mm emitting regions (rather than the inner disc) and do not produce high-resolution infrared spectra, so it is difficult to draw detailed comparisons.
}

{The main result of \citet{Walsh:2015jr} is that the inner regions of M dwarf discs appear more carbon rich than their higher mass counterparts. Their M dwarf disc model shows \ce{C2H2}/HCN column density ratios that are an order of magnitude lower than the observations suggest \citep{Pascucci:2008ht,Pascucci:2009gda}. They also suggest that simple organic molecules such as \ce{C2H2} and HCN have greater column densities in M dwarf discs, an effect that is also visible in our models: Fig. \ref{fig:fluxratiosvsrtaper} shows that smaller discs in our model grid have greater HCN/CO flux ratios, which is supportive of this. A much more detailed investigation of the inner disc chemistry, and comparisons with \textit{Spitzer} data, will be the subject of a forthcoming paper. In this paper, we keep our focus on molecules in the outer disc, which are observable at sub-mm wavelengths.
}

The disc models (in terms of chemical networks and element abundances) have been shown to reproduce the observed sub-mm line ratios in T Tauri discs (Rab et al. in prep.). Thus it should be feasible to model the line fluxes of brown dwarf discs, under the assumption that we can straightforwardly modify the input parameters of a typical T Tauri model to apply to the case of a brown dwarf. In this case, we can reasonably match the models and observations together.

\section{Results and discussion}

We have modelled all of the brown dwarf disc models, and the T Tauri reference model, with a limited 100-species chemical network. We extend this by modelling the fiducial brown dwarf model and T Tauri reference model with a large, 235-species network after finding that there are some molecules significant to the HCN and \hcop{} chemistry that are missing in the smaller network. However, we stress that the limited network is  sufficient for modelling CO, which is chemically simple, stable and well-known. Our CO sub-mm line fluxes change by less than $10\%$ when enlarging the size of the chemical network. We compare line flux ratios for both our brown dwarf and T Tauri models to observations in the literature, to check how closely the brown dwarfs follow their counterparts.

\subsection{Line ratios and flux predictions of brown dwarf discs}
\label{sec:lineratiosandfluxes}

\begin{figure}[tbh]
\centering
\includegraphics[width=\linewidth]{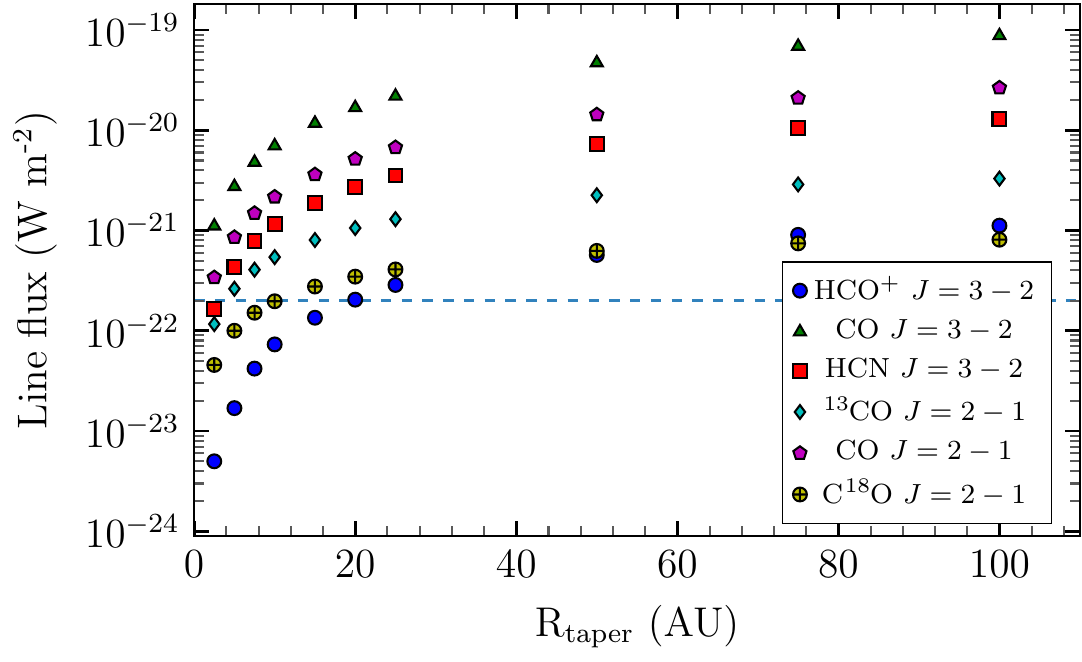}
\caption{Line flux predictions of the $\mdisc = 0.0004~\msun$ models. The dashed, horizontal line indicates a rough, pragmatic sensitivity limit for ALMA observations that is translated from a sensitivity of about $2.6~\mjy$ at $0.821~\kms$ spectral resolution, which is achievable in around 100 minutes (including overheads) with ALMA.}
\label{fig:fluxplot}
\end{figure}

To give an idea of the observability of common lines in the brown dwarf models, Fig. \ref{fig:fluxplot} shows line flux estimates across varying radii.

In the larger discs, we see that the line flux of some species becomes insensitive to the outer radius as the majority of the emission becomes optically thin. {In  brown dwarf discs with $\rtaper \lesssim 25 ~\au$, we find that} the line flux is insensitive to the gas mass because most of the CO emission is optically thick, which may hinder estimations of the disc gas mass. This effect can be seen in Fig. \ref{fig:fluxplot}, where the increase in line flux begins to flatten as the disc size increases beyond $\rtaper \gtrsim 25~\au$. However, there are no dramatic changes which suggest that the chemical processes discontinuously enter entirely different regimes when the disc is scaled beyond a certain threshold radius.

The \twco{} \jthrtwo{} emission is typically very optically thick (most emission comes from a line optical depth of $\tline \gtrsim 100$ in the fiducial model). The optical depth only approaches 1 at the very outer edges of the models, from where there is very little flux.  By contrast, most of the \hcn{} \jfrthr{} emission is from optical depths of $1 < \tline < 10$. The \hcop{} \jfrthr{} emission is very optically thin, with optical depths $\tau \approx 0.02$ throughout the disc.

\begin{figure}[tbh]
\centering
\includegraphics[width=\linewidth]{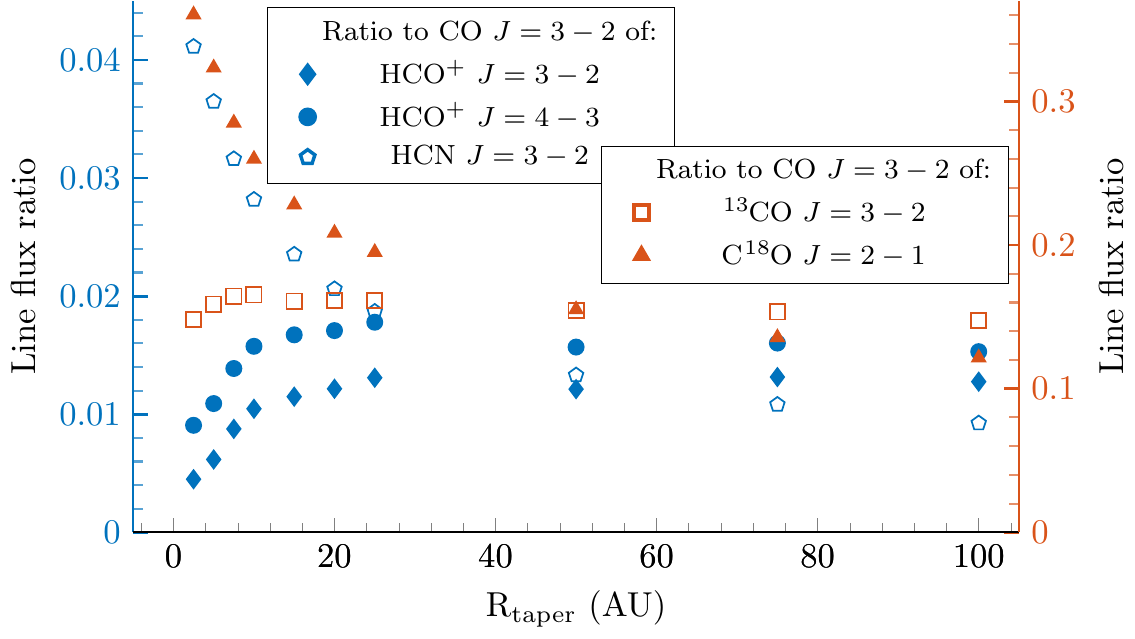}
\caption{Dependence of line flux ratios against \rtaper{}, for the line fluxes of a selection of molecules each divided by the \twco{} $J=3-2$ flux. The models are a subset of our model grid at $\mdisc=0.0004~\msun$. We note the differing scales on the colour-coded left- and right-hand ordinates. }
\label{fig:fluxratiosvsrtaper}
\end{figure}

{Chemically, gas-phase CO is  very stable} and is insensitive to changes in the chemical networks used. However, HCN and \hcop{} are molecules that are very sensitive to  the disc environment, the chemical network, and the available formation pathways. We find that in modelling the larger T Tauri disc, the small 100-species network used tends to {show higher HCN fluxes and lower \hcop{} fluxes relative to the large network}, as not all species that are relevant to the destruction and formation of these molecules are included in the small network (Rab et al., in prep.). Since we observe similar differences in flux ratios in the brown dwarf models, we extrapolate this and assume that the same holds for the brown dwarf regime.

Using the line ratios to interpret the HCN and \hcop{} line fluxes, we see that the HCN fluxes of the small network models are relatively high compared to our T Tauri models, and observations (though the observed HCN fluxes vary greatly). However, the HCN fluxes of the large network models are more comparable to the median observed T Tauri ratio. The small chemical network models appear to over-predict the HCN fluxes.

For \hcop{}, the small chemical network model appears to underpredict the line flux ratios. \hcop{} is a molecule that is very sensitive to the radiation environment, so some of the differences between the \hcop{} line ratios for the brown dwarf and T Tauri models may be accounted for by the less harsh radiation environment (that is, fewer X-rays) of the brown dwarf disc.  Our modelled line ratios show that \hcop{} lines in brown dwarf discs appear weaker than T Tauri discs, and that \hcn{} lines are fairly comparable.

Figure \ref{fig:fluxratiosvsrtaper} shows the dependence of line flux ratios on the taper radius of the brown dwarf disc. HCN and \hcop{} show opposite trends, where HCN flux ratios drop as the disc becomes larger and \hcop{} flux ratios increase. \thco{} flux ratios are fairly constant across the whole sample of radii, while \cego{} flux ratios drop significantly as the disc becomes larger. Small discs appear to have more extreme properties than larger discs ($\rtaper \gtrsim 30~\au$), where the line flux ratios stay fairly constant as \rtaper{} increases.

It is clear that large and complex chemical networks are needed to accurately model the lines from the more sensitive molecules. Because the line fluxes of \hcop{} and HCN in the small network models can differ by a factor of a few compared to the large network, the biases that the small 100-species chemistry network introduces should be accounted for in any interpretations of the sensitive molecules in these models. Future line flux observations of brown dwarfs -- comparing their flux ratios with that of our model predictions -- will either cement or disprove this for these key sub-mm lines. Observations of these line ratios will be a litmus test for the overall chemistry of brown dwarfs in comparison to discs around more massive stars.

\subsection{Disc geometry}

Because many brown dwarf discs are too small to be resolved, even with ALMA, the peak separation of lines such as the CO rotational lines can act as a discriminant in order to find discs that are very compact. 
 We expect that in a population of brown dwarf discs, the very compact discs will generally be distinguishable with spectrally-resolved observations, where the CO peak separation will be at least a few $\kms$.

Large peak separations ($\gtrsim 2~ \kms$) in CO sub-mm lines can only be produced by a small disc of $\rtaper \lesssim 5~\au$, thus enabling some very compact discs to be identified without knowledge of their inclination. If the disc is spatially resolved and its inclination and outer radius can be estimated from the spatial information, the shapes of molecular lines will then tell us about the radial distribution of the gas in the disc.

The modelled peak separations show a systematic dependence on disc mass. However, the disc mass dependence diminishes for smaller discs in comparison to the effects of changing \rout{}. Very compact brown dwarf discs of radius $\rout \lesssim 10~\au$ should be clearly identifiable by large peak separations of $\vsep \gtrsim 2.0~\kms$ in the \twco{} \jtwoone{} line.

Not changed here is the degree of disc flaring. Numerous near-infrared observations (for example, \citealp{Guieu:2007dj,AlvesdeOliveira:2013df,Mohanty:2004io}) do strongly suggest that most brown dwarf discs can be fitted with flaring indices $1 \leq \beta \leq 1.25$. However, good statistical constraints on the  flaring indices have not yet been formulated. Disc flaring may substantially affect the irradiation of the disc, and especially in a large disc, any robust interpretations of the line fluxes are dependent on having constraints on the flaring exponent $\beta$. Flaring is difficult to isolate without spatial information, because the SED is highly degenerate. However, infrared data such as the forbidden $\left[\mathrm{O}~\textsc{i}\right]~63~\micron$ line can help to break the degeneracies. ProDiMo models by \cite{Woitke:2010bya} of larger T Tauri discs show that across a wide range of disc structure parameters, there is an increase (sometimes of an order of magnitude or more) in $[\mathrm{O}~\textsc{i}]~63~\micron$ flux, as the flaring parameter $\beta$ increases from $1.0$ to $1.2$.

We did test whether the flaring parameter of our fiducial model produces results that are consistent with those of the higher mass T Tauri disc models. This was done by modifying the fiducial model to have different flaring exponents (leaving all other parameters the same), and we find that the $[\mathrm{O}~\textsc{i}]~63~\micron$ flux of a modified fiducial model with a flaring exponent of $\beta = 1.2$ is 20 times brighter than that of the same disc with $\beta = 1.0$. Thus, $[\mathrm{O}~\textsc{i}]~63~\micron$ observations are expected to provide strong constraints on disc flaring even in smaller brown dwarf discs.

\subsection{HCN chemistry}
\label{sec:hcnchemistry}

\begin{figure*}
\centering
\includegraphics[width=0.32\linewidth]{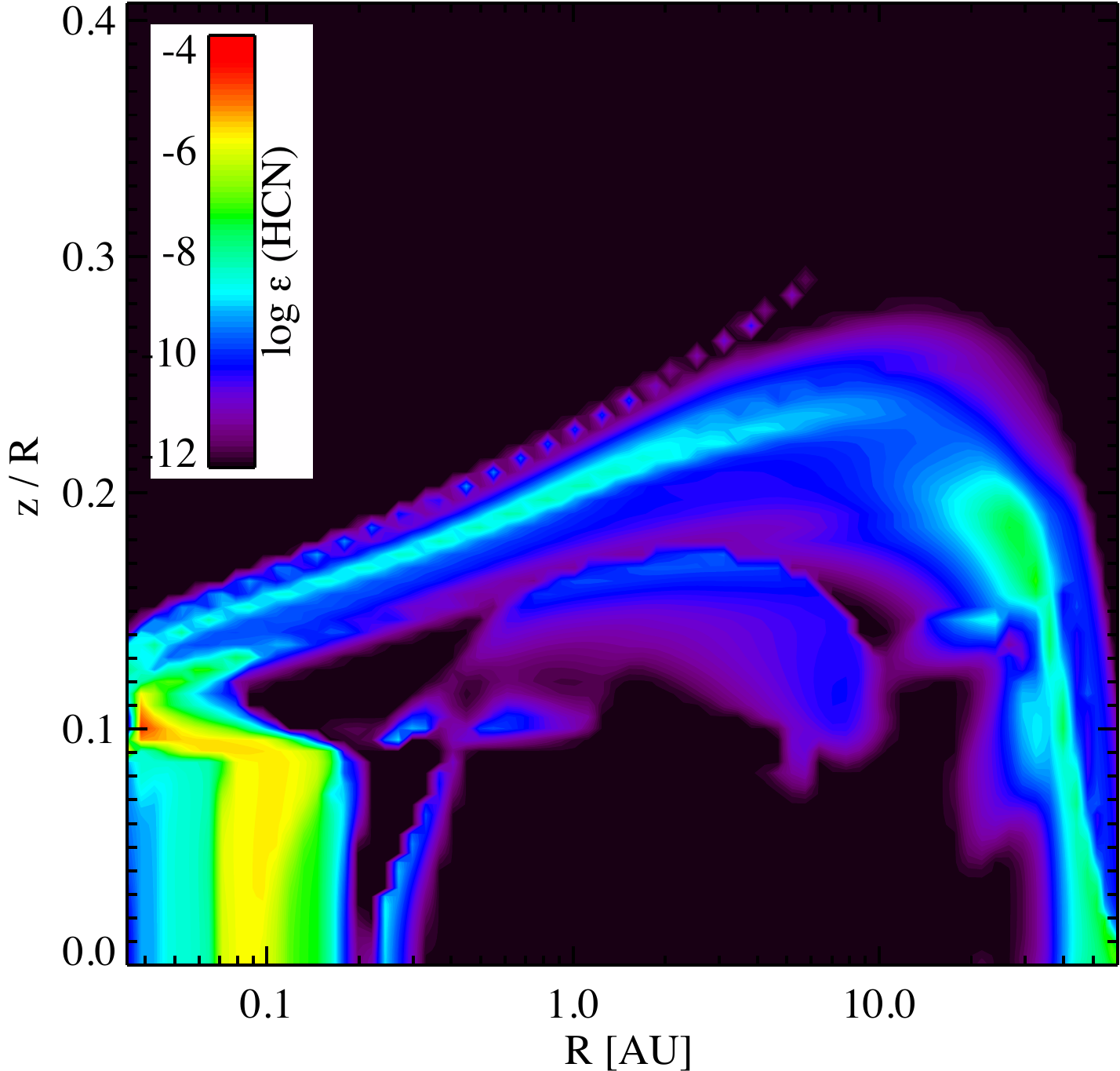}
\hspace{0.5em}
\includegraphics[width=0.32\linewidth]{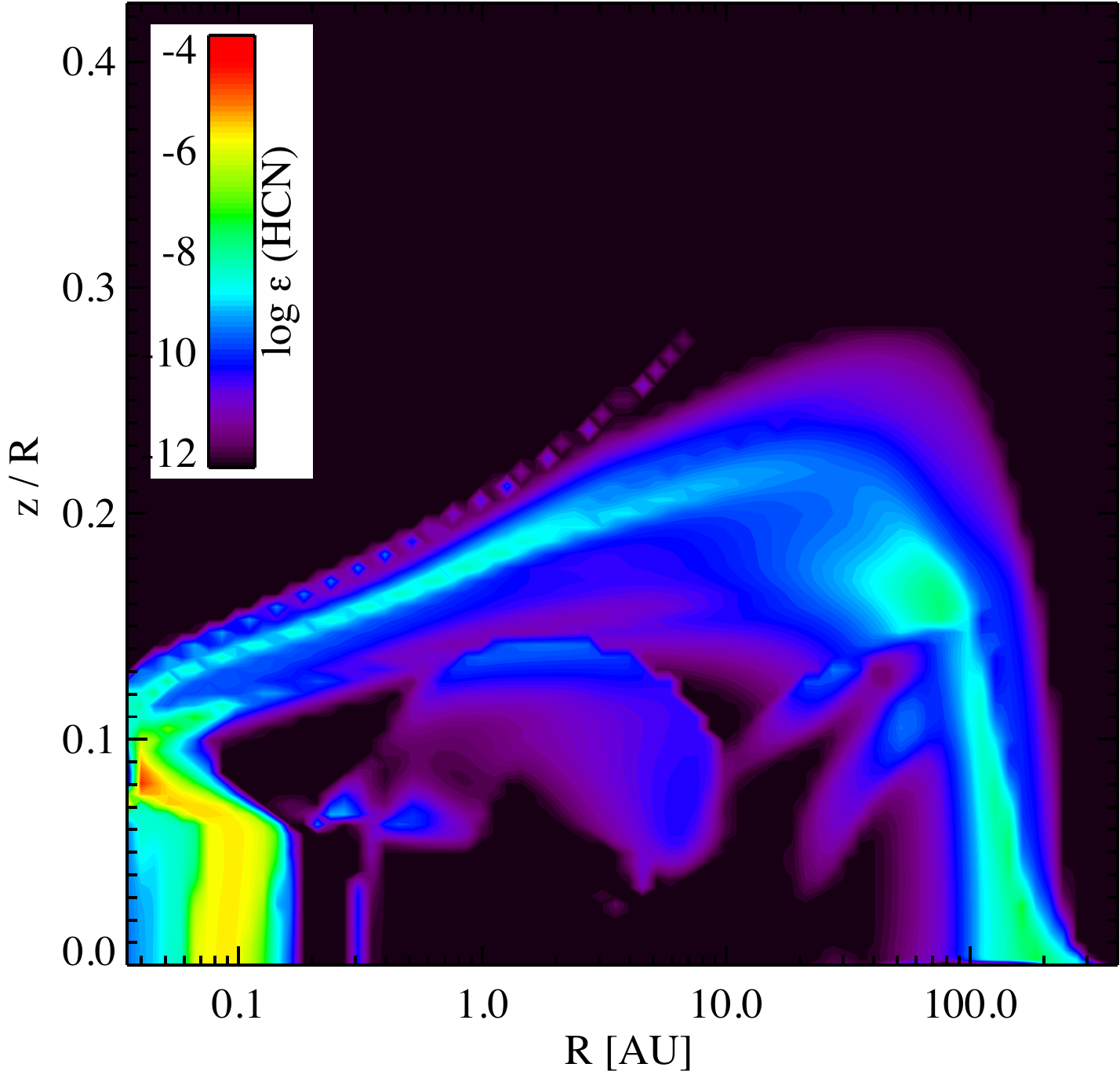}
\hspace{0.5em}
\includegraphics[width=0.32\linewidth]{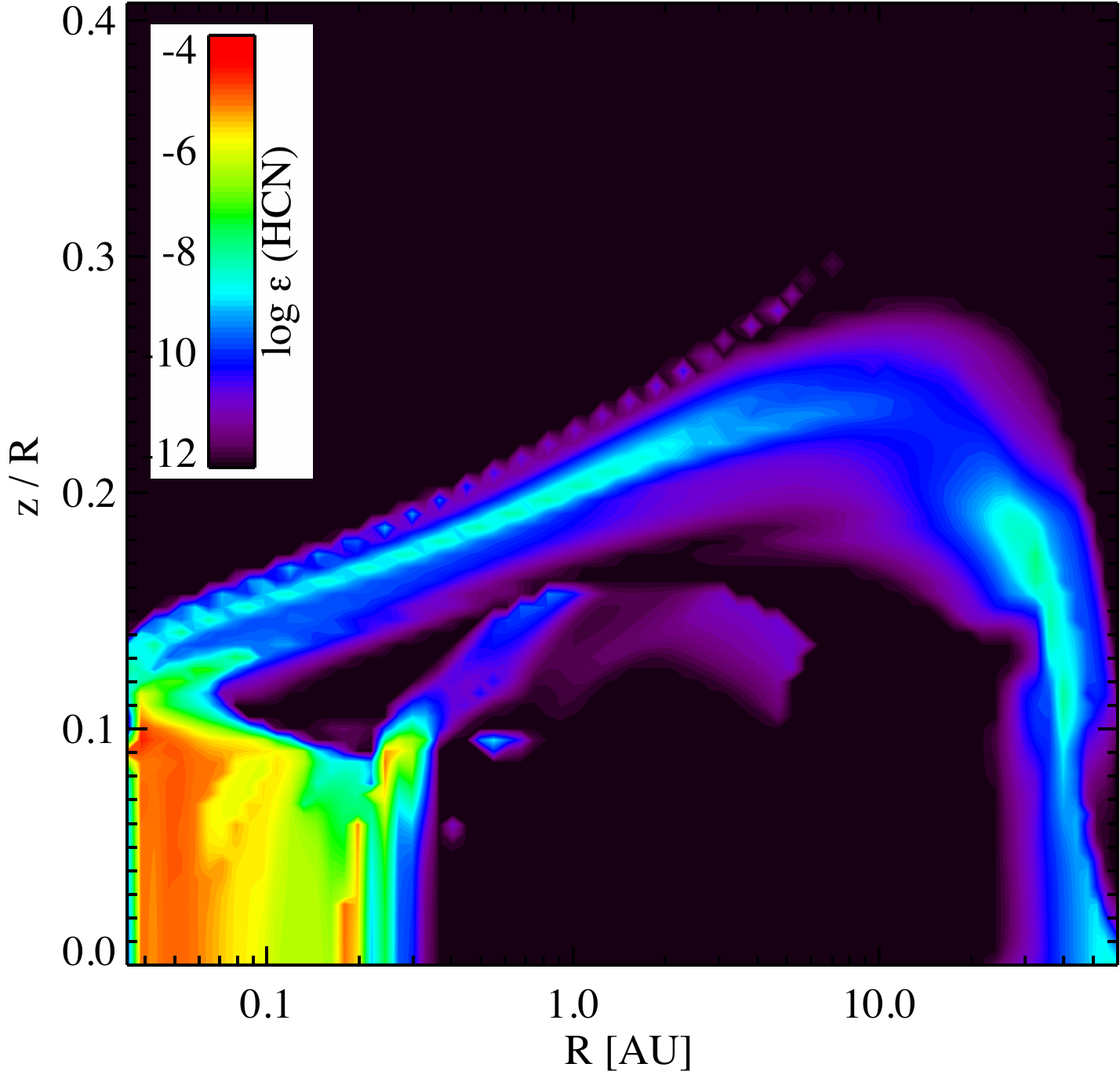}
\caption{
\textbf{Left, middle:} HCN gas abundances for two models where $\mdisc = 4 \times 10^{-4} ~\msun$, and $\rout=60~\au$ (the fiducial model) and $400~\au$.
\textbf{Right:} HCN gas abundances for the 235-species version of the fiducial model.}
\label{fig:hcnplots}
\end{figure*}

The HCN \jfrthr{} ($845.7~\micron$) and \jthrtwo{} ($1127.5~\micron$) lines trace layers of warm gas that are sensitive to the UV radiation field \citep{Aikawa:1999ty}, and the line is typically more optically thin than CO. As potentially one of the most readily observable molecules in brown dwarf discs, the capabilities of HCN to diagnose basic disc properties are of significant interest (to date, in the mm and sub-mm wavelengths, only \twco{} and dust continuum data have been published).

HCN has previously been found to be quite sensitive to the UV environment \citep{vanZadelhoff:2002kx}, and that it is easily destroyed by photodissociation. In stars with relatively weak levels of UV flux -- including brown dwarfs -- one might expect to see relatively bright HCN lines. However, this is not shown by our thermochemical models, where with the inclusion of a large chemical network we see HCN flux ratios that are comparable to observations of T Tauri discs (see Fig. \ref{fig:linefluxratios}).

\begin{figure*}[tbh]
\centering
\includegraphics[width=0.9\linewidth]{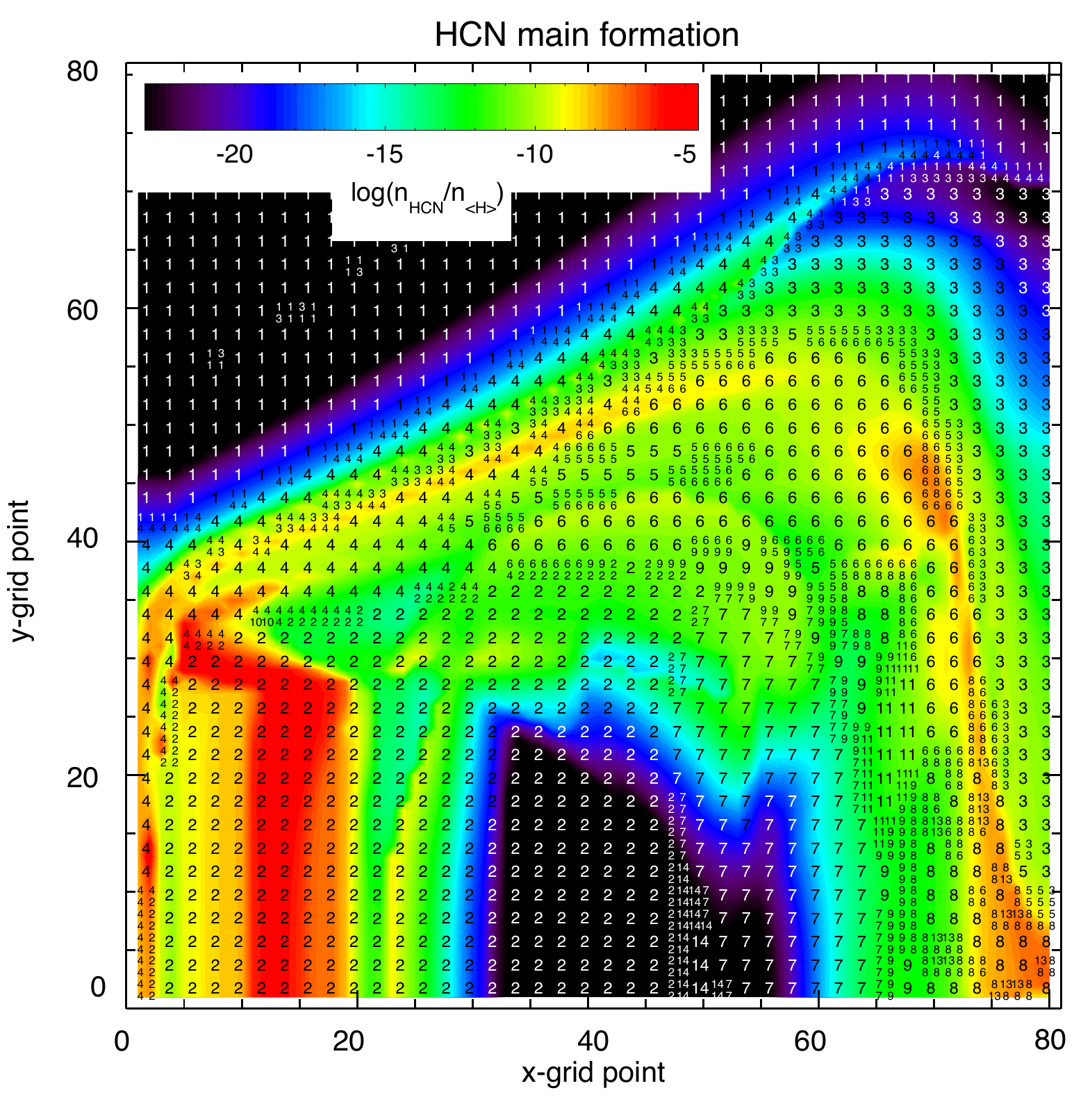}
\caption{Colour map of the HCN abundance of the fiducial model, overplotted with numbers that show the dominant formation reaction at each grid point. These reactions are enumerated in Table \ref{tab:hcnreactions}. For legibility, the grid points have been binned where the same reaction number is dominant in a square of four cells.
}
\label{fig:HCNformation_fiducial}
\end{figure*}

Fig. \ref{fig:hcnplots} shows the structure of HCN in both the fiducial model and a large brown dwarf model, where the HCN regions that are observable trace the warm upper layers of the gas in the disc. The chemical structures are very self-similar, the smaller model strongly resembling the larger model (with only the abscissa scale changed). The figure also shows the 235-species version of the fiducial model. We note that there is negligible sub-mm flux from the region of enhanced HCN abundance in the inner disc of the 235-species model.  Fig. \ref{fig:HCNformation_fiducial} shows which HCN reaction is most dominant in each cell of the 2D model, where the enumerated reaction numbers are explained in Table \ref{tab:hcnreactions}. 

Which reaction dominates the HCN chemistry depends on the complexity of the chemical network used. The model grid and fiducial model utilize a small, 100-species chemical network for reasons of minimizing computation time. However, modifying the fiducial model to use a larger chemical network with 235 species does introduce some changes. In the small network fiducial model, Fig. \ref{fig:HCNformation_fiducial} shows that in the outer parts of the disc where HCN is abundant, the formation is dominated by two neutral-neutral reactions: \ce{C      + NH2              \rightharpoonup HCN    + H} (reaction 6) and \ce{N      + CH3              \rightharpoonup HCN    + H2} (reaction 8). The warm, inner columns that harbour HCN all the way down to the midplane are dominated by the formation reaction \ce{H2     + CN               \rightharpoonup HCN    + H } (reaction 4).

The picture changes when modifying the fiducial model to include the advanced chemical network with 235 species, which shows that a simpler network is insufficient to accurately model HCN in a brown dwarf disc. That is, while a network with an incomplete selection of species and reactions is sufficient to model insensitive molecules such as CO, the small networks run the risk of omitting formation and destruction pathways that are very important to the more sensitive molecules.

While the same reactions dominate the warm surface strata and inner midplane regions of formation when using both the large and small chemical networks, when using the large network the outer disc is instead dominated by \ce{H      + H2CN             \rightharpoonup HCN    + H2}. The overall HCN formation rates and fluxes are lower in the small chemistry models because the introduction of \ce{H2CN} in the large chemical network plays a dominant role in both the formation and destruction of HCN. The destruction of HCN via the three-body reaction \ce{H + HCN + M \rightharpoonup H2CN + M}, where M is some {third body}, is a very efficient process in the outer disc that ends up decreasing the HCN abundances.

\begin{table}[tb]
\begin{center}
\caption[]{List of HCN formation and destruction reactions, to accompany Fig. \ref{fig:HCNformation_fiducial}. The abbreviations are as follows: (exc) denotes an excited state, $\gamma$(ISM) denotes {a UV photon}, $\gamma$(CR) denotes a cosmic ray-induced UV photon, and \ce{p(cosmic)} denotes a cosmic ray proton. We note that some of the literature which uses UMIST reactions denotes the latter three as PHOTON, CRPHOT, and CRP respectively.
}
\label{tab:hcnreactions}
\medskip
\begin{tabular}{l l r}
Formation reactions                                      \\ \midrule
       1  &  \ce{H2exc  + CN               \rightharpoonup HCN    + H       }\\
       2  &  \ce{HCN ice   + dust             \rightharpoonup HCN    + dust    } \\
       3  &  \ce{H      + HCN+             \rightharpoonup HCN    + H+      } \\
       4  &  \ce{H2     + CN               \rightharpoonup HCN    + H       } \\
       5  &  \ce{HCNH+  + e-               \rightharpoonup HCN    + H       } \\
       6  &  \ce{C      + NH2              \rightharpoonup HCN    + H       } \\
       7  &  \ce{HCN ice   + $\gamma$(ISM)           \rightharpoonup HCN           } \\
       8  &  \ce{N      + CH3              \rightharpoonup HCN    + H2      } \\
       9  &  \ce{CH     + NO               \rightharpoonup HCN    + O       } \\
      10  &  \ce{NH3    + HCNH+            \rightharpoonup HCN    + NH4+    } \\
      11  &  \ce{NH3    + CN               \rightharpoonup HCN    + NH2     } \\
      12  &  \ce{N      + CH2              \rightharpoonup HCN    + H       } \\
      13  &  \ce{N      + CH3              \rightharpoonup HCN    + H   + H    } \\
      14  &  \ce{HCN ice   + p(cosmic)              \rightharpoonup HCN            } \\
      15  &  \ce{H-     + CN               \rightharpoonup HCN    + e-      } \\
      16  &  \ce{Na     + HCNH+            \rightharpoonup HCN    + Na+   + H  }   \\
      17  &  \ce{OH     + CN               \rightharpoonup HCN    + O       } \\
      18  &  \ce{CH2    + NO               \rightharpoonup HCN    + OH      }     \\
      19  &  \ce{N      + HCO              \rightharpoonup HCN    + O       }    \\

Destruction reactions                                        \\ \midrule
       1  &  \ce{ H+     + HCN              \rightharpoonup HCN+   + H       } \\
       2  &  \ce{ HCN    + dust             \rightharpoonup HCN ice   + dust    } \\
       3  &  \ce{ HCN    + $\gamma$(ISM)           \rightharpoonup CN     + H       } \\
       4  &  \ce{ H3+    + HCN              \rightharpoonup HCNH+  + H2      } \\
       5  &  \ce{ N+     + HCN              \rightharpoonup HCN+   + N       } \\
       6  &  \ce{ H3O+   + HCN              \rightharpoonup HCNH+  + H2O     } \\
       7  &  \ce{ H      + HCN              \rightharpoonup CN     + H2      } \\
       8  &  \ce{ HCN    + HCO+             \rightharpoonup HCNH+  + CO      } \\
       9  &  \ce{ CH5+   + HCN              \rightharpoonup HCNH+  + CH4     } \\
      10  &  \ce{ He+    + HCN              \rightharpoonup CN+    + He      } \\
      11  &  \ce{ OH+    + HCN              \rightharpoonup HCNH+  + O       } \\
      12  &  \ce{ NH+    + HCN              \rightharpoonup HCNH+  + N       } \\
       13  &   \ce{    CH+    + HCN              \rightharpoonup HCNH+  + C       } \\
        14 &     \ce{  O      + HCN              \rightharpoonup CN     + OH      } \\

\end{tabular}
\end{center}

\end{table}

\subsection{HCO$^\mathit{+}$chemistry}
\label{sec:hcopchemistry}

Relative to the CO line fluxes, \hcop{} shows considerably lower flux in the brown dwarf models than in T Tauri discs (see Fig. \ref{fig:linefluxratios}). Dissociative recombination is the dominant form of \hcop{} destruction (\ce{HCO+ +  e- \rightharpoonup CO + H}) in most of the disc, including the regions where it is most abundant. This is the case for both the large and small chemical networks. The formation of \hcop{} is dominated almost singularly by ion-neutral reactions, however the pathways that dominate depend on the network used.

Combined with being relatively optically thin, \hcop{}  emission is a useful tracer of the ionization state of the disc. The gas temperature of the \hcop{} line emitting regions is warmer in the fiducial brown dwarf model than in the T Tauri model.

\begin{figure}[tbh]
\centering
\includegraphics[width=\linewidth]{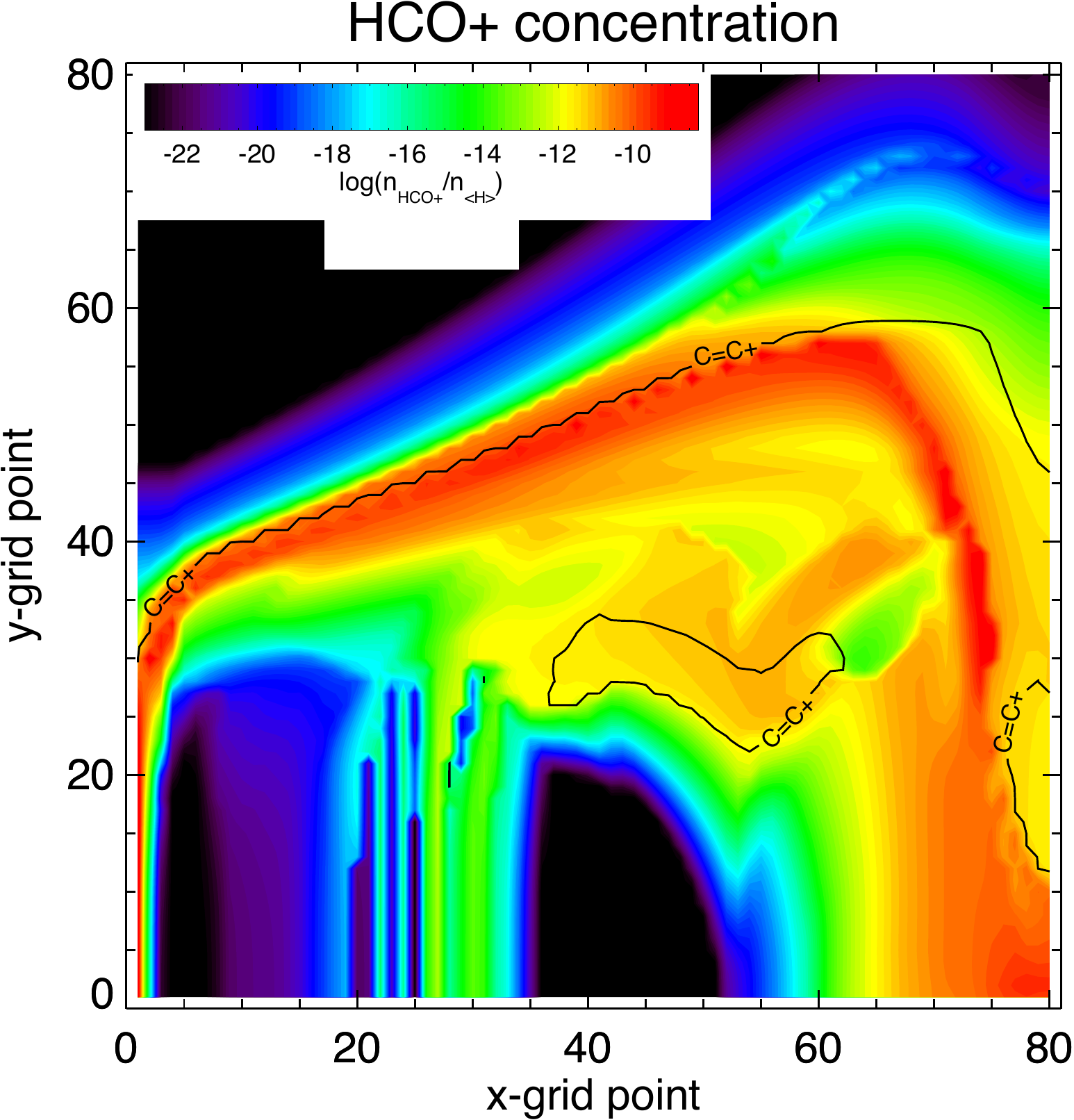}
\caption{\hcop{} concentration of the fiducial model using the large 235-species chemical network, showing where \hcop{} becomes much more abundant around the carbon ionization front.}
\label{fig:hcopchemistry}
\end{figure}

Fig. \ref{fig:hcopchemistry} shows that there are two main regions of \hcop{} formation: thin layers below the \ce{C \leftrightharpoons C+} ionization front through a formation path with \ce{H2} excited by UV radiation, and around the \ce{CO ice \leftrightharpoons CO gas} transition through the classical cosmic-ray driven path via \ce{H3+} (\citealp{Graedel:1982hf,Sternberg:1995dh}, equation 205). {The regions of \hcop{} in both the dense inner parts of the disc and the sparse outer layers are highly sensitive to X-ray and UV radiation respectively}, thus making it an excellent tracer molecule for observations.

ProDiMo outputs detailed formation and destruction information for each included molecule at each grid point in the model. To demonstrate the efficiency of \hcop{} formation in BD discs, we find that {around the CO ice line}, $76\%$ of the time that \ce{H2} is hit by an {X-ray photon} (\ce{H2 + $\gamma$ \rightharpoonup H2+ + e- + e-}), an \hcop{} molecule is produced. The \hcop{} chemistry in this region appears very smoothly resolved.

If we remove the UV excess from the fiducial model, \hcop{} becomes more abundant and the \jfrthr{} flux increases by a factor of about 30. However, decreasing the UV excess by only a factor of 10 relative to the fiducial model results in an increase in \jfrthr{} flux of only $9\%$. {The UV radiation efficiently ionizes atomic metals, and the resulting electrons efficiently destroy \hcop{} \citep{Teague:2015jk}.} The increase in flux is not so dramatic for longer wavelength lines. This could explain why the CO to \hcop{} flux ratios appear lower in T Tauri disc models than in brown dwarf disc models. {UV} radiation also works to boost the abundance of free electrons, which aids in the destruction of \hcop{} and ultimately works to suppress the \hcop{} abundances.

Around the carbon ionization transition, about one \hcop{} is created for every hundred \ce{H2} molecules that are excited by a UV photon. Notably, the \hcop{} abundance can drop by up to two orders of magnitude across one grid point (see Fig. \ref{fig:hcopchemistry}); the chemistry of this transition appears to be unresolved. Even when increasing the grid size of the model from $80\times 80$ points to $150 \times 150$, there is little visual difference between the models and there remain steep transitions in the \hcop{} abundance that may result in imprecise modelling of the \hcop{} line flux. However, despite the finer resolution there is no significant change ($\lesssim 1\%$) in the \twco{}\jtwoone{} and \hcop{} \jthrtwo{} fluxes. Some form of adaptive mesh refinement or a locally-defined increase in resolution may be useful, but if nearly doubling the (linear) resolution of the grid resulted in no significant change, it is likely that the smaller grid resolution is sufficient for analysing \hcop{} in brown dwarf discs.

\section{Conclusions}

We have produced a small grid of brown dwarf disc models which agrees with the observations of the brown dwarf $\rho$ Oph 102, while keeping the disc setup as T Tauri-like as possible. We thus test the hypothesis that we can model the structure and chemistry of brown dwarf discs in a broadly similar fashion to T Tauri discs.

Brown dwarf disc models appear subject to the same caveats that affect their larger T Tauri counterparts. When modelling more complex and sensitive species such as HCN and \hcop{}, it is necessary to use complex chemical networks because the dominant chemical pathways appear to change significantly. It is also important to ensure that regions with sharp chemical transitions are adequately resolved: this becomes even more important when modelling discs with gaps and discontinuities. However, once these effects are taken into account, we predict that the flux ratios of \hcop{} and HCN to CO for brown dwarf discs should be very similar to that of T Tauri discs. Observations of these lines will provide crucial tests of the models.

Very small disc models appear as would be expected for a compact Keplerian disc, where the peak separation of mm and sub-mm lines is of the order of a few \kms{}. The ability to easily identify the size of the gas disc despite the weakness of the lines -- without spatially resolving them -- may prove useful in a survey capacity. The inclination may also be estimated from spatially resolved dust continuum data. However, analysing disc chemistry is better done in discs that are of the size of our fiducial model or larger: here, ALMA has the resolution and sensitivity to resolve the sub-mm dust disc (and perhaps in the \twco{} \jthrtwo{} line), and the sub-mm line profiles of CO, \hcop{}, and HCN are likely to be detectable.

\section{Acknowledgements}

We would like to thank the anonymous referee for the comments that improved the manuscript, and the Center for Information Technology of the University of Groningen for providing access to the Peregrine high performance computing cluster. This publication makes use of VOSA, developed under the Spanish Virtual Observatory project supported from the Spanish MICINN through grant AyA2011-24052. RCH acknowledges funding by the Austrian Science Fund (FWF): project number P24790.

\bibliographystyle{aa} 
\bibliography{mylocalbib,fullbib_local}

\begin{thebibliography}{69}
\expandafter\ifx\csname natexlab\endcsname\relax\def\natexlab#1{#1}\fi

\bibitem[{Aikawa \& Herbst(1999)}]{Aikawa:1999ty}
Aikawa, Y. \& Herbst, E. 1999, A{\&}A, 351, 233

\bibitem[{Alves~de Oliveira {et~al.}(2013)Alves~de Oliveira, {\'A}brah{\'a}m,
  Marton, Pinte, Kiss, Kun, K{\'o}sp{\'a}l, Andre, \&
  K{\"o}nyves}]{AlvesdeOliveira:2013df}
Alves~de Oliveira, C., {\'A}brah{\'a}m, P., Marton, G., {et~al.} 2013, A{\&}A,
  559, 126

\bibitem[{Antonellini(2016)}]{stefanothesis}
Antonellini, S. 2016, PhD thesis, Rijksuniversiteit Groningen

\bibitem[{Aresu {et~al.}(2011)Aresu, Kamp, Meijerink, Woitke, Thi, \&
  Spaans}]{Aresu:2011cm}
Aresu, G., Kamp, I., Meijerink, R., {et~al.} 2011, A{\&}A, 526, A163

\bibitem[{Baron {et~al.}(2003)Baron, Hauschildt, Allard, Lentz, Aufdenberg,
  Schweitzer, \& Barman}]{Baron:2003tk}
Baron, E., Hauschildt, P.~H., Allard, F., {et~al.} 2003, Proceedings of the
  International Astronomical Union, 210, 19

\bibitem[{Basri(1998)}]{Basri:1998tg}
Basri, G. 1998, IAU Colloq. 193: Variable Stars in the Local Group, 134, 394

\bibitem[{Bate(2009)}]{Bate:2009br}
Bate, M.~R. 2009, MNRAS, 392, 590

\bibitem[{Bate(2012)}]{Bate:2012hy}
Bate, M.~R. 2012, MNRAS, 419, 3115

\bibitem[{Bate {et~al.}(2003)Bate, Bonnell, \& Bromm}]{Bate:2003cv}
Bate, M.~R., Bonnell, I.~A., \& Bromm, V. 2003, MNRAS, 339, 577

\bibitem[{Bayo {et~al.}(2008)Bayo, Rodrigo, Barrado~y Navascu{\'e}s, Solano,
  Guti{\'e}rrez, Morales-Calderon, \& Allard}]{Bayo:2008bk}
Bayo, A., Rodrigo, C., Barrado~y Navascu{\'e}s, D., {et~al.} 2008, A{\&}A, 492,
  277

\bibitem[{Burrows {et~al.}(1997)Burrows, Marley, Hubbard, Lunine, Guillot,
  Saumon, Freedman, Sudarsky, \& Sharp}]{Burrows:1997ua}
Burrows, A., Marley, M., Hubbard, W.~B., {et~al.} 1997, ApJ, 491, 856

\bibitem[{Burrows {et~al.}(1995)Burrows, Saumon, Guillot, Hubbard, \&
  Lunine}]{Burrows:1995gr}
Burrows, A., Saumon, D., Guillot, T., Hubbard, W.~B., \& Lunine, J.~I. 1995,
  Nature, 375, 299

\bibitem[{{DENIS Consortium}(2005)}]{Consortium:2005vp}
{DENIS Consortium}. 2005, VizieR Online Data Catalog, 2263

\bibitem[{Dorschner {et~al.}(1995)Dorschner, Begemann, Henning, Jaeger, \&
  Mutschke}]{Dorschner:1995wq}
Dorschner, J., Begemann, B., Henning, T., Jaeger, C., \& Mutschke, H. 1995,
  A{\&}A, 300, 503

\bibitem[{Draine(1978)}]{Draine:1978ec}
Draine, B.~T. 1978, ApJS, 36, 595

\bibitem[{Dubrulle {et~al.}(1995)Dubrulle, Morfill, \&
  Sterzik}]{Dubrulle:1995jn}
Dubrulle, B., Morfill, G., \& Sterzik, M. 1995, Icarus, 114, 237

\bibitem[{Evans {et~al.}(2009)Evans, Dunham, J{\o}rgensen, Enoch, Mer{\'\i}n,
  van Dishoeck, Alcal{\'a}, Myers, Stapelfeldt, Huard, Allen, Harvey, van
  Kempen, Blake, Koerner, Mundy, Padgett, \& Sargent}]{Evans:2009bk}
Evans, N.~J., Dunham, M.~M., J{\o}rgensen, J.~K., {et~al.} 2009, ApJS, 181, 321

\bibitem[{Evans {et~al.}(2003)Evans, Allen, Blake, Boogert, Bourke, Harvey,
  Kessler, Koerner, Lee, Mundy, Myers, Padgett, Pontoppidan, Sargent,
  Stapelfeldt, van Dishoeck, Young, \& Young}]{Evans:2003bo}
Evans, N. J.~I., Allen, L.~E., Blake, G.~A., {et~al.} 2003, Publications of the
  Astronomical Society of the Pacific, 115, 965

\bibitem[{Gillon {et~al.}(2016)Gillon, Jehin, Lederer, Delrez, de~Wit,
  Burdanov, Van~Grootel, Burgasser, Triaud, Opitom, Demory, Sahu,
  Bardalez~Gagliuffi, Magain, \& Queloz}]{Gillon:2016hl}
Gillon, M., Jehin, E., Lederer, S.~M., {et~al.} 2016, Nature, 533, 221

\bibitem[{Graedel {et~al.}(1982)Graedel, Langer, \& Frerking}]{Graedel:1982hf}
Graedel, T.~E., Langer, W.~D., \& Frerking, M.~A. 1982, ApJS, 48, 321

\bibitem[{Guieu {et~al.}(2007)Guieu, Pinte, Monin, M{\'e}nard, Fukagawa,
  Padgett, Noriega-Crespo, Carey, Rebull, Huard, \& Guedel}]{Guieu:2007dj}
Guieu, S., Pinte, C., Monin, J.~L., {et~al.} 2007, A{\&}A, 465, 855

\bibitem[{Han {et~al.}(2013)Han, Jung, Udalski, Sumi, Gaudi, Gould, Bennett,
  Tsapras, Szyma{\'{n}}ski, Kubiak, Pietrzy{\'{n}}ski, Soszy{\'{n}}ski,
  Skowron, Koz{\l}owski, Poleski, Ulaczyk, Wyrzykowski, Pietrukowicz,
  Collaboration, Abe, Bond, Botzler, Chote, Freeman, Fukui, Furusawa, Harris,
  Itow, Ling, Masuda, Matsubara, Muraki, Ohnishi, Rattenbury, Saito, Sullivan,
  Sweatman, Suzuki, Tristram, Wada, {Yock, P. C. M.}, Collaboration, Batista,
  Christie, Choi, DePoy, Dong, Hwang, Kavka, Lee, Monard, Natusch, Ngan, Park,
  Pogge, Porritt, Shin, Tan, Yee, Collaboration, Alsubai, Bozza, Bramich,
  Browne, Dominik, Horne, Hundertmark, Ipatov, Kains, Liebig, Snodgrass,
  Steele, Street, \& Collaboration}]{Han:2013by}
Han, C., Jung, Y.~K., Udalski, A., {et~al.} 2013, ApJ, 778, 38

\bibitem[{Harvey {et~al.}(2012)Harvey, Henning, Liu, M{\'e}nard, Pinte, Wolf,
  Cieza, Evans, \& Pascucci}]{Harvey:2012cz}
Harvey, P.~M., Henning, T., Liu, Y., {et~al.} 2012, ApJ, 755, 67

\bibitem[{Hauschildt \& Baron(1999)}]{Hauschildt:1999vw}
Hauschildt, P.~H. \& Baron, E. 1999, Journal of Computational and Applied
  Mathematics, 109, 41

\bibitem[{Helling {et~al.}(2008)Helling, Ackerman, Allard, Dehn, Hauschildt,
  Homeier, Lodders, Marley, Rietmeijer, Tsuji, \& Woitke}]{Helling:2008hp}
Helling, C., Ackerman, A., Allard, F., {et~al.} 2008, MNRAS, 391, 1854

\bibitem[{Helling \& Casewell(2014)}]{Helling:2014fx}
Helling, C. \& Casewell, S. 2014, Astron Astrophys Rev, 22, 80

\bibitem[{Helling \& Woitke(2006)}]{Helling:2006gp}
Helling, C. \& Woitke, P. 2006, A{\&}A, 455, 325

\bibitem[{Hollenbach {et~al.}(2008)Hollenbach, Kaufman, Bergin, \&
  Melnick}]{Hollenbach:2008ho}
Hollenbach, D., Kaufman, M.~J., Bergin, E.~A., \& Melnick, G.~J. 2008, ApJ,
  690, 1497

\bibitem[{Johnson {et~al.}(2010)Johnson, Aller, Howard, \&
  Crepp}]{Johnson:2010gu}
Johnson, J.~A., Aller, K.~M., Howard, A.~W., \& Crepp, J.~R. 2010, Publications
  of the Astronomical Society of the Pacific, 122, 905

\bibitem[{Kamp {et~al.}(2013)Kamp, Thi, Meeus, Woitke, Pinte, Meijerink,
  Spaans, Pascucci, Aresu, \& Dent}]{Kamp:2013ix}
Kamp, I., Thi, W.~F., Meeus, G., {et~al.} 2013, A{\&}A, 559, A24

\bibitem[{Kamp {et~al.}(2010)Kamp, Tilling, Woitke, Thi, \&
  Hogerheijde}]{Kamp:2010ek}
Kamp, I., Tilling, I., Woitke, P., Thi, W.~F., \& Hogerheijde, M. 2010, A{\&}A,
  510, A18

\bibitem[{Lawrence {et~al.}(2007)Lawrence, Warren, Almaini, Edge, Hambly,
  Jameson, Lucas, Casali, Adamson, Dye, Emerson, Foucaud, Hewett, Hirst,
  Hodgkin, Irwin, Lodieu, McMahon, Simpson, Smail, Mortlock, \&
  Folger}]{Lawrence:2007hu}
Lawrence, A., Warren, S.~J., Almaini, O., {et~al.} 2007, MNRAS, 379, 1599

\bibitem[{Manara {et~al.}(2015)Manara, Testi, Natta, \&
  Alcal{\'a}}]{Manara:2015ey}
Manara, C.~F., Testi, L., Natta, A., \& Alcal{\'a}, J.~M. 2015, A{\&}A, 579,
  A66

\bibitem[{McElroy {et~al.}(2013)McElroy, Walsh, Markwick, Cordiner, Smith, \&
  Millar}]{McElroy:2013ki}
McElroy, D., Walsh, C., Markwick, A.~J., {et~al.} 2013, A{\&}A, 550, A36

\bibitem[{Min {et~al.}(2005)Min, Hovenier, \& de~Koter}]{Min:2005uy}
Min, M., Hovenier, J.~W., \& de~Koter, A. 2005, A{\&}A

\bibitem[{Min {et~al.}(2016)Min, Rab, Woitke, Dominik, \&
  M{\'e}nard}]{Min:2016hr}
Min, M., Rab, C., Woitke, P., Dominik, C., \& M{\'e}nard, F. 2016, A{\&}A, 585,
  A13

\bibitem[{Mohanty {et~al.}(2013)Mohanty, Greaves, Mortlock, Pascucci, Scholz,
  Thompson, Apai, Lodato, \& Looper}]{Mohanty:2013kl}
Mohanty, S., Greaves, J., Mortlock, D., {et~al.} 2013, ApJ, 773, 168

\bibitem[{Mohanty {et~al.}(2004)Mohanty, Jayawardhana, Natta, Fujiyoshi,
  Tamura, \& Barrado~y Navascu{\'e}s}]{Mohanty:2004io}
Mohanty, S., Jayawardhana, R., Natta, A., {et~al.} 2004, ApJ, 609, L33

\bibitem[{Mu{\v z}i{\'c} {et~al.}(2012)Mu{\v z}i{\'c}, Scholz, Geers,
  Jayawardhana, \& Tamura}]{Muzic:2012jo}
Mu{\v z}i{\'c}, K., Scholz, A., Geers, V., Jayawardhana, R., \& Tamura, M.
  2012, ApJ, 744, 134

\bibitem[{Natta {et~al.}(2002)Natta, Testi, Comer~n, Oliva, D'Antona, Baffa,
  Comoretto, \& Gennari}]{Natta:2002ea}
Natta, A., Testi, L., Comer~n, F., {et~al.} 2002, A{\&}A, 393, 597

\bibitem[{{\"O}berg {et~al.}(2010){\"O}berg, Qi, Fogel, Bergin, Andrews,
  Espaillat, van Kempen, Wilner, \& Pascucci}]{Oberg:2010kf}
{\"O}berg, K.~I., Qi, C., Fogel, J. K.~J., {et~al.} 2010, ApJ, 720, 480

\bibitem[{{\"O}berg {et~al.}(2011){\"O}berg, Qi, Fogel, Bergin, Andrews,
  Espaillat, Wilner, Pascucci, \& Kastner}]{Oberg:2011ew}
{\"O}berg, K.~I., Qi, C., Fogel, J. K.~J., {et~al.} 2011, ApJ, 734, 98

\bibitem[{Ochsenbein {et~al.}(2000)Ochsenbein, Bauer, \&
  Marcout}]{Ochsenbein:2000fm}
Ochsenbein, F., Bauer, P., \& Marcout, J. 2000, Astronomy and Astrophysics
  Supplement Series, 143, 23

\bibitem[{Oppenheimer {et~al.}(2000)Oppenheimer, Kulkarni, \&
  Stauffer}]{Oppenheimer:2000vo}
Oppenheimer, B.~R., Kulkarni, S.~R., \& Stauffer, J.~R. 2000, Protostars and
  Planets IV, 1313

\bibitem[{Pascucci {et~al.}(2008)Pascucci, Apai, Hardegree-Ullman, Kim, Meyer,
  \& Bouwman}]{Pascucci:2008ht}
Pascucci, I., Apai, D., Hardegree-Ullman, E.~E., {et~al.} 2008, ApJ, 673, 477

\bibitem[{Pascucci {et~al.}(2009)Pascucci, Apai, Luhman, Henning, Bouwman,
  Meyer, Lahuis, \& Natta}]{Pascucci:2009gda}
Pascucci, I., Apai, D., Luhman, K., {et~al.} 2009, ApJ, 696, 143

\bibitem[{Pascucci {et~al.}(2013)Pascucci, Herczeg, \& Carr}]{Pascucci:2013te}
Pascucci, I., Herczeg, G., \& Carr, J.~S. 2013, ApJ

\bibitem[{Qi {et~al.}(2003)Qi, Kessler, Koerner, Sargent, \& Blake}]{Qi:2003dn}
Qi, C., Kessler, J.~E., Koerner, D.~W., Sargent, A.~I., \& Blake, G.~A. 2003,
  ApJ, 597, 986

\bibitem[{Ricci {et~al.}(2012)Ricci, Testi, Natta, Scholz, \&
  de~Gregorio-Monsalvo}]{2012ApJ...761L..20R}
Ricci, L., Testi, L., Natta, A., Scholz, A., \& de~Gregorio-Monsalvo, I. 2012,
  ApJ, 761, L20

\bibitem[{Ricci {et~al.}(2014)Ricci, Testi, Natta, Scholz,
  de~Gregorio-Monsalvo, \& Isella}]{Ricci:2014im}
Ricci, L., Testi, L., Natta, A., {et~al.} 2014, ApJ, 791, 20

\bibitem[{Salter {et~al.}(2011)Salter, Hogerheijde, van~der Burg, Kristensen,
  \& Brinch}]{Salter:2011jn}
Salter, D.~M., Hogerheijde, M.~R., van~der Burg, R. F.~J., Kristensen, L.~E.,
  \& Brinch, C. 2011, A{\&}A, 536, 80

\bibitem[{Skrutskie {et~al.}(2006)Skrutskie, Cutri, Stiening, Weinberg,
  Schneider, Carpenter, Beichman, Capps, Chester, Elias, Huchra, Liebert,
  Lonsdale, Monet, Price, Seitzer, Jarrett, Kirkpatrick, Gizis, Howard, Evans,
  Fowler, Fullmer, Hurt, Light, Kopan, Marsh, McCallon, Tam, Van~Dyk, \&
  Wheelock}]{Skrutskie:2006hl}
Skrutskie, M.~F., Cutri, R.~M., Stiening, R., {et~al.} 2006, The Astronomical
  Journal, 131, 1163

\bibitem[{Sternberg \& Dalgarno(1995)}]{Sternberg:1995dh}
Sternberg, A. \& Dalgarno, A. 1995, ApJS, 99, 565

\bibitem[{Teague {et~al.}(2015)Teague, Semenov, Guilloteau, Henning, Dutrey,
  Wakelam, Chapillon, \& Pietu}]{Teague:2015jk}
Teague, R., Semenov, D., Guilloteau, S., {et~al.} 2015, A{\&}A, 574, A137

\bibitem[{Testi {et~al.}(2016)Testi, Natta, Scholz, Tazzari, Ricci, \&
  de~Gregorio-Monsalvo}]{Testi:2016tw}
Testi, L., Natta, A., Scholz, A., {et~al.} 2016, A{\&}A, 593, A111

\bibitem[{Tilling {et~al.}(2012)Tilling, Woitke, Meeus, Mora, Montesinos,
  Riviere-Marichalar, Eiroa, Thi, Isella, Roberge, Martin-Zaidi, Kamp, Pinte,
  Sandell, Vacca, M{\'e}nard, Mendigut{\'\i}a, Duch{\^e}ne, Dent, Aresu,
  Meijerink, \& Spaans}]{Tilling:2012cj}
Tilling, I., Woitke, P., Meeus, G., {et~al.} 2012, A{\&}A, 538, 20

\bibitem[{van~der Plas {et~al.}(2016)van~der Plas, M{\'e}nard, Ward-Duong,
  Bulger, Harvey, Pinte, Patience, Hales, \& Casassus}]{vanderPlas:2016do}
van~der Plas, G., M{\'e}nard, F., Ward-Duong, K., {et~al.} 2016, ApJ, 819, 102

\bibitem[{van Zadelhoff {et~al.}(2002)van Zadelhoff, Aikawa, Hogerheijde, \&
  Van~Dishoeck}]{vanZadelhoff:2002kx}
van Zadelhoff, G.~J., Aikawa, Y., Hogerheijde, M.~R., \& Van~Dishoeck, E.~F.
  2002, A{\&}A, 397, 789

\bibitem[{Walsh {et~al.}(2015)Walsh, Nomura, \& van Dishoeck}]{Walsh:2015jr}
Walsh, C., Nomura, H., \& van Dishoeck, E. 2015, A{\&}A, 582, A88

\bibitem[{Wiebe {et~al.}(2008)Wiebe, Semenov, \& Henning}]{Wiebe:2008dy}
Wiebe, D.~S., Semenov, D.~A., \& Henning, T. 2008, Astron. Rep., 52, 941

\bibitem[{Witte {et~al.}(2011)Witte, Helling, Barman, Heidrich, \&
  Hauschildt}]{Witte:2011kn}
Witte, S., Helling, C., Barman, T., Heidrich, N., \& Hauschildt, P.~H. 2011,
  A{\&}A, 529, A44

\bibitem[{Witte {et~al.}(2009)Witte, Helling, \& Hauschildt}]{Witte:2009br}
Witte, S., Helling, C., \& Hauschildt, P.~H. 2009, A{\&}A, 506, 1367

\bibitem[{Woitke \& Helling(2003)}]{Woitke:2003cs}
Woitke, P. \& Helling, C. 2003, A{\&}A, 399, 297

\bibitem[{Woitke \& Helling(2004)}]{Woitke:2004ie}
Woitke, P. \& Helling, C. 2004, A{\&}A, 414, 335

\bibitem[{Woitke {et~al.}(2009)Woitke, Kamp, \& Thi}]{Woitke:2009jf}
Woitke, P., Kamp, I., \& Thi, W.~F. 2009, A{\&}A, 501, 383

\bibitem[{Woitke {et~al.}(2016)Woitke, Min, Pinte, Thi, Kamp, Rab, Anthonioz,
  Antonellini, Baldovin-Saavedra, Carmona, Dominik, Dionatos, Greaves,
  G{\"u}del, Ilee, Liebhart, M{\'e}nard, Rigon, Waters, Aresu, Meijerink, \&
  Spaans}]{Woitke:2016gp}
Woitke, P., Min, M., Pinte, C., {et~al.} 2016, A{\&}A, 586, A103

\bibitem[{Woitke {et~al.}(2010)Woitke, Pinte, Tilling, M{\'e}nard, Kamp, Thi,
  Duch{\^e}ne, \& Augereau}]{Woitke:2010bya}
Woitke, P., Pinte, C., Tilling, I., {et~al.} 2010, MNRAS, 405, L26

\bibitem[{Woitke {et~al.}(2011)Woitke, Riaz, Duch{\^e}ne, Pascucci, Lyo, Dent,
  Phillips, Thi, M{\'e}nard, Herczeg, Bergin, Brown, Mora, Kamp, Aresu,
  Brittain, de~Gregorio-Monsalvo, \& Sandell}]{Woitke:2011fo}
Woitke, P., Riaz, B., Duch{\^e}ne, G., {et~al.} 2011, A{\&}A, 534, A44

\bibitem[{Zubko {et~al.}(1996)Zubko, Mennella, Colangeli, \&
  Bussoletti}]{Zubko:1996fn}
Zubko, V.~G., Mennella, V., Colangeli, L., \& Bussoletti, E. 1996, MNRAS, 282,
  1321

\end{thebibliography}

\end{document}